\documentclass[aps,showpacs,nofootinbib,amssymb,preprint]{revtex4}
\usepackage{graphicx,epsfig}
\usepackage{bm}
\usepackage{amsmath, amsfonts, amssymb}
\usepackage{color}
\newcommand{\nn}{\nonumber}
\newcommand{\be}{\begin{equation}}
\newcommand{\ee}{\end{equation}}
\newcommand{\bea}{\begin{eqnarray}}
\newcommand{\eea}{\end{eqnarray}}
\def\al{\alpha}

\def\siml{{\ \lower-1.2pt\vbox{\hbox{\rlap{$<$}\lower6pt\vbox{\hbox{$\sim$}}}}\ }} 
\def\lQ{\Lambda_{\rm QCD}}

\newcommand{\MS}{\overline{\rm MS}}

\newcommand{\ord}{{\cal O}}

\def\msb{{\overline{\rm MS}}}

\preprint{ \vbox{
\hbox{UWThPh-2011-17}
}}

\begin{document}

\title{\boldmath 
The static hybrid potential in $D$ dimensions at short distances
\unboldmath}
\author{Antonio Pineda$^1$ and Maximilian Stahlhofen$^{1,2}$}
\affiliation{$^1$Grup de F\'\i sica Te\`orica, Universitat
Aut\`onoma de Barcelona, E-08193 Bellaterra, Barcelona, Spain\\
$^2$University of Vienna, Faculty of Physics, Boltzmanngasse 5, A-1090 Wien, Austria \vspace{1 cm}}

\date{\today \vspace{1 cm}}

\begin{abstract}
\noindent
We compute the energy of a static hybrid, i.e. of a hybrid quarkonium with static quark and antiquark, at short distances in $D=4,3$ dimensions.
The soft contribution to this energy is the static potential of a color octet quark-antiquark pair at short distances, which is known at two loops for arbitrary $D$. We have checked this expression employing thermal field theory methods. 
Using the effective field theory pNRQCD we calculate the ultrasoft contributions to the hybrid (and singlet) static energy at the two-loop level. 
We then present new results for the static hybrid energy/potential and the hybrid decay width in three and four dimensions.
Finally we comment on the meaning of the perturbative results in two space-time dimensions, where the hybrid does not exist.
\end{abstract}
\maketitle

\section{Introduction}

In a recent paper~\cite{Pineda:2010mb}
we have studied the potential and energy of a static color-singlet quark-antiquark state at short distances in $D$ space-time dimensions (though with special emphasis on the three-dimensional case). It is the aim of this 
paper to perform a similar analysis for the static hybrid energy and the associated octet potential. 

The energies of static hybrid systems are very interesting quantities. They can teach us a 
lot about the dynamics of QCD, and can be potentially relevant for the theoretical description 
of physical hybrids made of heavy quarks. Their behavior at long distances might contain information on the 
detailed dynamics responsible for confinement, whether it is due to strings and, if so, of which kind. 
Most relevant for us, however, is that the hybrid energy at short distances approximately equals the sum of the static octet potential energy
and the respective gluelump mass. This relation was first quantified in Refs.~\cite{Jorysz:1987qj,FM} and recently exposed in an unified and 
model-independent framework~\cite{Brambilla:1999xf,Bali:2003jq} using the effective field theory ``potential nonrelativistic QCD'' (pNRQCD)~\cite{Pineda:1997bj} (for a review see \cite{Brambilla:2004jw}).

The physics of the static hybrid system in the short distance limit is governed by, at least, two physical scales.
One is the soft scale $\sim 1/r$, the inverse distance between quark and antiquark, the other is the ultrasoft scale $\sim \Delta V \equiv V_o - V_s$, where $V_o$ and $V_s$ are the static octet and singlet potential, respectively. The effective theory pNRQCD is particularly suitable to study this limit, as it profits from the large scale separation: $1/r \gg \Delta V$.
For small distance $r$ the static hybrid state can be understood to consist of a color octet quark-antiquark state (acting as 
a static source in the octet representation) coupled to some ultrasoft and/or non-perturbative gluons to form an overall color singlet state.
A close relative of the static hybrid (as we will see below) is the gluelump, which consists of a static octet source (of whatever origin) attached to nonperturbative gluons forming a color-singlet state.

In order to gain a deeper understanding of the physics of the static hybrid system one can also consider how it is qualitatively affected by changing the number of dimensions from four (4D) to three (3D) or two (2D). The three-dimensional result is also important on its own. Four dimensional thermal QCD 
effectively undergoes a dimensional reduction for large temperatures. 
Therefore, determining the renormalization group (RG) structure for the static potential in 
three dimensions might open the way to a resummation of logarithms at finite temperature. 
Three dimensional space-time is moreover a good testing ground for renormalon issues, 
since the linear power divergences associated with renormalons in four dimensions become logarithmic divergences 
in three dimensions and can be traced back using dimensional regularization (see Ref.~\cite{Pineda:2010mb}). 
Last but not least the computations in less than four dimensions represent important consistency checks of the theoretical framework used to describe the 4D hybrid (and singlet) systems at short distances.

The color-octet static potential is obtained by integrating out the soft scale and has received quite some attention 
in the last decade. For $D=4$, it was computed with two-loop precision in Ref.~\cite{Kniehl:2004rk}, the leading three-loop logarithms were obtained in Ref.~\cite{short}, 
and the associated resumation of logarithms was carried out in Ref.~\cite{RG}.  
The 4D two-loop result of Ref.~\cite{Kniehl:2004rk} was confirmed in Ref.~\cite{Brambilla:2010xn} by computing the correlator of two Polyakov loops at finite temperature, which is then related to the singlet and octet static potentials. 
From the terms given in Ref.~\cite{Kniehl:2004rk} we have also been able to read off the general $D$ dimensional expression for the static octet potential, and verified it using the method developed in Ref.~\cite{Brambilla:2010xn}. This expression is infrared divergent for $D=3$. The divergences must be canceled by the ultraviolet divergences of the ultrasoft contribution. We calculate the latter in pNRQCD for arbitrary $D$ up to two loops and present the results in full detail on a diagram-by-diagram basis. We also compute the $D$ dimensional ultrasoft two-loop correction to the static singlet energy providing an independent confirmation of the results obtained in Refs.~\cite{Eidemuller:1997bb,Brambilla:2006wp}. The one-loop ultrasoft singlet and octet contributions are basically identical except for trivial modifications. The ultrasoft two-loop octet computation has previously been considered in Ref.~\cite{Brambilla:2009bi}, where a one-to-one correspondence to the singlet result in the temporal ($A_0=0$) gauge was claimed.
This statement can be rather problematic due to the singular nature of the gluon propagator in this gauge. Actually, for $D=4$, the anomalous dimension of the octet static potential we find differs from the one obtained using the relation between singlet and octet computation based on the $A_0=0$ gauge argument of Ref.~\cite{Brambilla:2009bi}. In three dimensions the problem is even more acute, as the result obtained from the suggested singlet-octet correspondence does not have the correct divergence structure to make the computation renormalizable and to derive a meaningful anomalous dimension. Our explicit ultrasoft two-loop result in Feynman gauge, in contrast, does, and exactly cancels the infrared divergences of the static octet potential for $D=3$. This represents a strong cross-check of our calculation.

The discussion in the present paper closely follows the lines of Ref.~\cite{Pineda:2010mb}. Therefore, we will skip some details by
referring to that reference. The outline of the paper is as follows. In Sec.~\ref{pNRQCD} we introduce the theoretical setup and present the bare two-loop soft and ultrasoft results for the color octet. The expressions for the relevant pNRQCD diagrams in Feynman gauge are shown in the Appendix, both for the color octet and singlet case. In Sec.~\ref{secD4} we obtain the static hybrid energy at next-to-next-to-next-to-leading logarithmic (N$^3$LL) order (apart from the soft three-loop matching constant) for $D=4$. Sec.~\ref{sec3D} contains the complete RG improved ultrasoft calculation up to NNLL level in three dimensions. In Sec.~\ref{secD2} we discuss the situation in two dimensions, and in Sec.~\ref{conclusion} we present our conclusions.

\section{pNRQCD}
\label{pNRQCD}

Up to the next-to-leading order (NLO) in the multipole expansion (and irrespectively of the space-time dimension) 
the effective Lagrangian density of pNRQCD in the static limit takes the form\footnote{In principle, to fully account for the nonperturbative contributions at $\ord(r^2)$ we have to add the term
\begin{align}
\delta {\cal L}
=
  g \frac{V_C (r)}{8} r^ir^j{\rm Tr} \left\{  {\rm O}^\dagger \left[{\bf D}^i {\bf E}^j , {\rm O}\right] \right\}
\label{lagVC}
\end{align}
to the Lagrangian in Eq.~\eqref{pnrqcd0}.}:
\begin{align}
&  {\cal L} =
{\rm Tr} \Biggl\{ {\rm S}^\dagger \left( i\partial_0  - V_s(r)   \right) {\rm S} 
 + {\rm O}^\dagger \left( iD_0 - V_o(r)  \right) {\rm O} \Biggr\} 
\nonumber\\
& \qquad
+ g V_A ( r) {\rm Tr} \left\{  {\rm O}^\dagger {\bf r} \cdot {\bf E} \,{\rm S}
+ {\rm S}^\dagger {\bf r} \cdot {\bf E} \,{\rm O} \right\} 
  + g \frac {V_B (r)}{2} {\rm Tr} \left\{  {\rm O}^\dagger \left\{{\bf r} \cdot {\bf E} , {\rm O}\right\}\right\} +  \ord(r^2)\,.
\label{pnrqcd0}
\end{align}
We define color singlet and octet fields for the quark-antiquark system by $S = S({\bf r},{\bf R},t)$ and 
$O^a =  O^a({\bf r},{\bf R},t)$ respectively. ${\bf R} \equiv ({\bf x}_1+{\bf x}_2)/2$ is the center position of the system.
In order for $S$ and $O^a$ to have the proper free-field normalization in color space they are related to the fields in Eq.~\eqref{pnrqcd0} as follows:
\begin{align}
{\rm S} \equiv \frac{ 1\!\!{\rm l}_c }{ \sqrt{N_c}} S\,, \qquad {\rm O} \equiv  \frac{ T^a }{ \sqrt{T_F}}O^a, 
\label{norm}
\end{align}
where $T_F=1/2$ for the fundamental representation of $SU(N_c)$.
All gluon fields in Eq.~\eqref{pnrqcd0} are evaluated 
in ${\bf R}$ and the time $t$, in particular the chromoelectric field ${\bf E} \equiv {\bf E}({\bf R},t)$ and the ultrasoft covariant derivative
$iD_0 {\rm O} \equiv i \partial_0 {\rm O} - g [A_0({\bf R},t),{\rm O}]$.

In the following we will use the index ``$B$'' to explicitly denote bare quantities. Parameters without this index are understood to be renormalized. 
We will furthermore use the notation $D\equiv 4+2\epsilon \equiv n+2\epsilon_n$, where $\epsilon_n=\frac{D-n}{2}$ 
parameterizes the (typically infinitesimal) difference to the closest integer dimension $n=4$, 3, 2. 
The bare parameters of the theory are the coupling constant $\al_B$ ($g_B$) and the potentials $V_{\{s,o,A,B\}, B}(r)$, generically denoted by $V_B$. The associated renormalized coefficients $\al(\nu)$ and $V_{\{s,o,A,B\}}(r; \nu)$ are the Wilson coefficients of the effective Lagrangian and depend on the renormalization scale ($\nu$). They are typically fixed at a scale smaller than (or similar to) $1/r$ and larger than the ultrasoft and any other scale in the problem by matching the effective to the underlying theory, which in this case is QCD in the static limit.

In our convention 
$\al_B$ has integer mass dimension, $[\al_B]=[\al]=M^{4-n}$ ($[g_B^2]=M^{4-D}$), and is related to $g_B$ by
\be
\alpha_B=\frac{g_B^2\nu^{2\epsilon_n}}{4\pi}\,,
\ee
where $\nu$ is the renormalization scale.
It has a special status since 
it does not receive corrections from other Wilson coefficients of the 
effective theory. Therefore, 
it can be renormalized multiplicatively:
\be
\al_B=Z_{\al}\al
\,,
\ee
where
\be
Z_{\al}
=1+
\sum_{s=1}^{\infty}Z^{(s)}_{\al}\frac{1}{\epsilon_n^s}\,.
\ee
The RG equation of $\al$ is
\be
\nu\frac{d}{d\nu}\al\equiv \al\beta(\al;\epsilon_n)=2\epsilon_n\al+\al\beta(\al;0)\,.
\ee
In the limit $\epsilon_n \rightarrow 0$
\be
\nu\frac{d}{d\nu}\al\equiv \al\beta(\al;0)
\equiv \al\beta(\al)=-2\al \frac{d}{d\al}Z^{(1)}_{\al}\,.
\ee

For the octet potential we employ an additive renormalization convention: 
\be
\label{VBsplitting}
V_{o,B}=V_o+\delta V_o\,.
\ee
The counterterm $\delta V_o$ generally depends on the Wilson coefficients of the effective theory, i.e. on $\al$ and $V$, 
and on the number of space-time dimensions.
Using the minimal subtraction (MS) renormalization scheme in $D(n)$ dimensions we define
\be
\delta V_o
=
\sum_{s=1}^{\infty}Z^{(s)}_{V_o}\frac{1}{\epsilon_n^s}\,.
\ee
From the scale independence of the octet bare potential
\be
\nu \frac{d}{d\nu}V_{o,B}=0
\,,
\ee
one obtains the RG equation of $V_o$. It can schematically be written as:
\be
\nu \frac{d}{d\nu}V_o=B(V)\,, \label{VRGE}
\ee
\be
B(V)\equiv -\left(\nu \frac{d}{d\nu}\delta V_o\right).
\ee
Note that $B(V)$ is, in general, a function of all the potentials appearing in the Lagrangian. 
Note as well that Eq.~\eqref{VRGE} implies that all the $1/\epsilon_n$ poles disappear once the derivative 
with respect to the renormalization scale is performed.
 This imposes some constraints on $\delta V_o$:
\bea
\label{Z1}
%\text{At }
{\cal O}(1/\epsilon_n): \qquad
&&B(V)=-2\al \frac{\partial}{\partial\al}Z^{(1)}_{V_o}\,,\; \label{1loopBV}
\\
%\text{at }
{\cal O}(1/\epsilon^2_n): \qquad
&&
\label{Z2}
B(V)\frac{\partial}{\partial V}Z^{(1)}_{V_o}
+
\al \beta (\al)\frac{\partial}{\partial\al}Z^{(1)}_{V_o}
+
2\al \frac{\partial}{\partial\al}Z^{(2)}_{V_o}=0\,,
\eea
and so on. 

\subsection{Bare results in $D$ dimensions}

This subsection summarizes the bare results of the calculations relevant in this work.
In $D$ dimensional momentum space we can write the bare octet potential as
\begin{align}
{\tilde V}_{o,B}=\frac{1}{2N_c}g_B^2\frac{1}{{\bf k}^2}
\sum_{n=0}^{\infty}g_B^{2n}{\bf k}^{2n\epsilon}\frac{\tilde c^{(o)}_n(D)}{(4\pi)^{nD/2}}
\,,
\label{momspacepot}
\end{align}
where $\tilde c^{(o)}_0(D)=1$ and $\tilde c^{(o)}_1(D)=\tilde c^{(s)}_1(D)$, $\tilde c^{(s)}_1(D)$ being 
the analogously defined one-loop coefficient for the static singlet potential, see Ref.~\cite{Schroder:1999sg}.
The two-loop coefficient $\tilde c^{(o)}_2(D)$ differs from $\tilde c^{(s)}_2(D)$, which was computed in Ref.~\cite{Schroder:1998vy}. 
We denote this difference by
\begin{align}
\delta \tilde c_2(D) = \tilde c^{(o)}_2(D) - \tilde c^{(s)}_2(D)
\,.
\end{align}
$\delta \tilde c_2(4)$ was first obtained in Ref.~\cite{Kniehl:2004rk}. The result was confirmed in Ref.~\cite{Brambilla:2010xn} using thermal QCD and effective theory methods to compute the correlator of two Polyakov loops and relate it to the singlet and octet potentials. Although no explicit expression is given one can also read the $D$ dimensional coefficient $\delta \tilde c_2(D)$ off the results in Ref.~\cite{Kniehl:2004rk}. Using the approach developed in Ref.~\cite{Brambilla:2010xn} we have been able to confirm that expression for arbitrary $D$.
It reads
\begin{align}
\delta \tilde c^{(o)}_2(D)
=C_A^2
\frac{\pi ^3  \Gamma (\epsilon \!+\! \frac{1}{2}) \sec ^2(\pi 
   \epsilon ) \left(\epsilon  (2 \epsilon +3) \Gamma (\epsilon \!-\! \frac{1}{2}) \Gamma (2 \epsilon \!+\!1) \csc (\pi  \epsilon )
 +\pi \Gamma (3 \epsilon \!+\!\frac{3}{2}) \sec (\pi  \epsilon) \right)}
 {\Gamma (\frac{3}{2} \!-\! \epsilon ) \Gamma (2 \epsilon \!+\!1)^2 \Gamma \left(3 \epsilon \!+\! \frac{1}{2}\right)}
\,.
\end{align}
 
After Fourier transformation to position space Eq.~\eqref{momspacepot} becomes
(see e.g. Ref.~\cite{Pascual:1984zb})
\bea
\label{VsBD}
V_{o,B}
&=&
\frac{1}{2N_c}g_B^2
\sum_{n=0}^{\infty}\frac{g_B^{2n}r^{-2(n+1)\epsilon}}{r}
\frac{\tilde c^{(o)}_{n}(D)}{(4\pi)^{nD/2}}
\frac{\Gamma[1/2+(n+1)\epsilon]}{2^{2-2n\epsilon}\pi^{3/2+\epsilon}\Gamma[1-n\epsilon]}
\\
\nn
&\equiv&
\frac{1}{2N_c}g_B^2\sum_{n=0}^{\infty}\frac{g_B^{2n}c^{(o)}_n(D)r^{-2(n+1)\epsilon}}{r}\,.
\eea

\medskip

We now consider the ultrasoft bare correction to the hybrid energy at ${\cal O}(r^2)$ in the multipole expansion.
It can be determined from the calculation of the bare octet self-energy $\Sigma_B^{us}(E)$ in terms of (1PI) loop diagrams in pNRQCD.
The octet propagator including the ultrasoft corrections (but neglecting nonperturbative ones) then takes the form
\begin{align}
\int dt e^{iEt}\langle O^a(t)O^b(0)\rangle \sim \frac{i}{E-V_o^B-\Sigma_B^{us}(E)}
\,.
\end{align}
To extract the ultrasoft correction to the static octet energy\footnote{This correction is understood to be complex as long as we do not explicitly disentangle the real part associated with the physical energy and the imaginary part associated with the decay width, see subsection~\ref{secObservables}.} at ${\cal O}(r^2)$ it is sufficient to set $E=V_o^B|_{\ord(r^0)}$ in the bare expression for the self-energy\footnote{In order to consistently treat the 3D infrared divergence at $\ord(r^0)$ (which is canceled by an ultraviolet divergence of nonperturbative origin) in dimensional regularization it is important to set $E=V_o^B$ before performing the loop integrations of the self-energy diagrams in this ``bare'' approach. As a check we have computed the corresponding off-shell diagrams and explicitly performed the renormalization of the potentials and the octet field. Finally taking the ``renormalized'' on-shell limit $E \to V_o$ gives the same result for the ultrasoft correction to the static hybrid energy.}, i.e. $\delta E_{o,B}^{us}=\Sigma_B^{us}(E=V_o^B) + \ord(r^3)$.

\begin{figure}[ht]
\includegraphics[width=0.25 \textwidth]{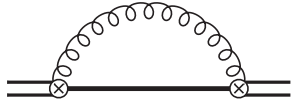}
%\makebox[0.0cm]{\phantom b}
%\epsfxsize=5truecm \epsfbox{usoctet3D.eps}
\caption{\it One-loop octet self-energy diagram at $\ord(r^2)$.}
\label{Oct1looprsq}
\end{figure}

The one-loop result equals the analogous singlet correction~\cite{Pineda:1997ie,short,KP1} except for a change of the color factor and the replacement $\Delta V \equiv V_o-V_s \to -\Delta V$:
\begin{align}
\delta E^{us}_{o,B}({\rm 1-loop})
=
-g^2\frac{1}{2N_c}V_A^2(1+\epsilon)
\frac{\Gamma[2+\epsilon]\Gamma[-3-2\epsilon]}{\pi^{2+\epsilon}}
{\bf r}^2\,(-\Delta V_B)^{3+2\epsilon}\,.
\label{USbare1loop}
\end{align}
In Feynman gauge it comes from the pNRQCD diagram shown in Fig.~\ref{Oct1looprsq}.
The two-loop expression is new and reads
\begin{align}
\label{USbare2loop}
\delta E^{us}_{s,B}({\rm 2-loop})
&=
g^4\frac{1}{2N_c}C_AV_A^2\Gamma[-3-4\epsilon]
\frac{1}{(2\pi)^2}\frac{1}{4\pi^{2+2\epsilon}}\Gamma^2[1+\epsilon]
\\
\nn
&
\times
\left[
g(\epsilon)-(1+2\epsilon)g_1(\epsilon)+g_{o}(\epsilon)
\right]
{\bf r}^2\,(-\Delta V_B)^{3+4\epsilon}\,,
\end{align}
where
\begin{align}
g(\epsilon)&=\frac{2 \epsilon^3+6 \epsilon^2+8 \epsilon+3}{\epsilon \left(2 \epsilon^2+5 \epsilon+3\right)}
-\frac{2 \epsilon \Gamma
   (-2 \epsilon-2) \Gamma (-2 \epsilon-1)}{(2 \epsilon+3) \Gamma (-4 \epsilon-3)}
\,,
\\
g_1(\epsilon)&=\frac{6 \epsilon^3+17 \epsilon^2+18 \epsilon+6}{\epsilon^2 
\left(2 \epsilon^2+5 \epsilon+3\right)}+\frac{4 (\epsilon+1)
   n_f T_f}{\epsilon (2 \epsilon+3) N_c}+\frac{2
   \left(\epsilon^2+\epsilon+1\right) \Gamma (-2 \epsilon-2) \Gamma (-2 \epsilon-1)}{\epsilon (2 \epsilon+3)
   \Gamma (-4 \epsilon-3)}
\,,
\\
g_o(\epsilon)&=\frac{2^{-4 \epsilon -5} (2 \epsilon +3) (2 \epsilon ^2+2 \epsilon +1)  \Gamma (-2 \epsilon -\frac{7}{2}) 
\Gamma^2 (-\epsilon -\frac{3}{2}) \Gamma (2 \epsilon +\frac{9}{2})}{\epsilon\,  \Gamma (-4 \epsilon -3) \Gamma^2 (\epsilon +1)}
\,.
\label{go}
\end{align}
It is the central result of this work. In the Appendix we separately present the analytic expressions for the two-loop self-energy diagrams involved in the calculation of Eq.~\eqref{USbare2loop} in the Feynman gauge. We have also computed the corresponding two-loop expressions for the color singlet, which provide an explicit confirmation of the result obtained in Refs.~\cite{Eidemuller:1997bb,Brambilla:2006wp} and can also be found in the Appendix.

%%%%%%%%%%%%%%%%%%%%%%%%%%%%%%%%%%
The computation of the ultrasoft octet self-energy has also been addressed in Ref.~\cite{Brambilla:2009bi}, 
where it was argued that the result could be related to the singlet one (which is gauge invariant) in the temporal ($A_0=0$) gauge. In order to 
reach this conclusion, the effects involving the gluonic content of the hybrid in the soft and ultrasoft computation were neglected (see Eq. 17 in Ref.~\cite{Brambilla:2009bi}). Then, after choosing the $A_0=0$ gauge, it is indeed easy to see that all non-vanishing self-energy diagrams at $\ord(r^2)$ are, apart from the color factor and the sign of $\Delta V$, the same for singlet and octet. Requiring gauge invariance of the final outcome one thus would obtain 
\begin{align}
V_A^2 \frac{T_F }{(D-1) (N_c^2-1)N_c} {\bf r}^2 \int_0^\infty \!\! dt\,  e^{t\, \Delta V_B} 
  \langle vac|
g{\bf E}_E^a(t) 
\phi(t,0)^{\rm adj}_{ab} g{\bf E}_E^b(0) |vac \rangle\,
\label{EoA0gauge}
\end{align}
for the bare ultrasoft contribution to the static octet energy (in Euclidean spacetime) at $\ord(r^2)$ and to all orders in $\al$.

At $\ord(\al)$ (one-loop) Eq.~\eqref{EoA0gauge} agrees with Eq.~\eqref{USbare1loop}. However, at $\ord(\al^2)$ (two-loop) Eqs.~\eqref{USbare2loop} and~\eqref{EoA0gauge} are different, since $g_o(\epsilon)=0$ in Eq. \eqref{EoA0gauge}. This is in contradiction with our non-zero result for $g_o(\epsilon)$ in Eq.~\eqref{go}, which was obtained by an explicit two-loop calculation in Feynman gauge (see App.~\ref{octdiagrams}). The existence of a non-zero value for $g_o$ is required on a 
theoretical basis in Sec.~\ref{sec3D}. Otherwise $\delta E^{us}_{o,B}$ would not produce the right divergence structure in $D=3$ dimensions to make 
the theory renormalizable nor lead to a consistent RG equation for the potential, as there would be some divergences proportional to logarithms of the ultrasoft scale left after the subtraction of the one-loop subdivergences. Our explicit computation, in contrast, is consistent with the renormalizability of the theory, and moreover perfectly cancels the infrared divergences of the soft computation. This represents a strong check of our computation. 
 In Sec.~\ref{secD4} we will see, that with our result in Eq.~\eqref{USbare2loop} we also obtain a different NLL anomalous dimension of the octet potential $V_o$ in four dimensions compared to Ref.~\cite{Brambilla:2009bi}. The corresponding cancelation of soft infrared and ultrasoft ultraviolet divergences in four dimensions can however not be checked, since the required soft four-loop result is not known at present.

Actually, the fact that there are problems with the naive usage of $A_0=0$ gauge becomes apparent at several points of the calculation. In pNRQCD this already happens at ${\cal O}(r^0)$ in the multipole expansion for $D=3$. At this order the bare soft result for $V_o$ is infrared divergent (cf. Eq.~\eqref{Z1Vo}), but in the $A_0=0$ gauge there are no ultrasoft octet self-energy diagrams to cancel that divergence.
It is also remarkable that in this gauge the space-like strings of the rectangular Wilson loop contribute to the static potential, whereas in Feynman gauge they are expected to be negligible\footnote{In Ref.~\cite{Schroder:1999sg} it was explicitly shown that the space-like strings in Feynman gauge do not contribute to the 4D static singlet potential through two loops.}. This can be easily visualized for $D=2$, since then time and space coordinates can be symmetrically interchanged (in the Euclidean), showing that in the $A_0=0$ gauge the contribution to the singlet potential comes exclusively from the space-like strings. Actually that is true for arbitrary $D$ at tree level. For a detailed analysis at one loop see e.g. Ref.~\cite{Hand:1995rg} and the references therein. Overall, we think it would be very interesting to quantitatively study the effect of the asymptotic gluonic degrees of freedom on the static singlet and hybrid energy in the $A_0=0$ gauge.

To the above discussion one should also add the $\ord(r^2)$ ultrasoft contributions from octet self-energy diagrams without singlet-octet transitions.
They are proportional to $V_B^2$ or $V_C$ and may have logarithmic ultraviolet divergences, relevant for the determination of the counterterm ($\delta V_o$) and the anomalous dimension of $V_o$.
These divergences can, however, not be proportional to positive powers of $\Delta V$, because the ultrasoft scale $\Delta V$ does not appear in the diagrams with onshell external fields, and the corresponding contributions vanish in dimensional regularization. 
Therefore they cannot solve the $A_0=0$ gauge problem described above, since the first discrepancies between our two-loop result in Eq.~\eqref{USbare2loop} and the $\ord(\al^2)$ term in Eq.~\eqref{EoA0gauge} are linear in $\Delta V$ for $D=3$ and cubic for $D=4$.

\subsection{Observables}
\label{secObservables}
At short distances we approximate the bare propagator of a generic hybrid system as follows:
\begin{align}
\int dt e^{i E t}\langle H^aO^a(t)H^bO^b(0)\rangle \sim \frac{i}{E - V_o^B - \Sigma_B(E)+i\epsilon}
\,.
\label{HybridProp}
\end{align}
$H^a$ stands for the gluonic content of the hybrid ($H^a=B^a, E^a$, etc.) and $\Sigma_B(E)$ is the self-energy of the system expressed in terms of bare parameters (potentials, couplings). 
It accounts for effects at the ultrasoft and the nonperturbative scale.

Let us now construct some observables. In this paper we focus on the hybrid static 
energy and decay width. 
Their definition can become gauge dependent for unstable particles.
Fortunately this is not so in our case, since both the on-shell and
pole definition (see e.g. Ref.~\cite{Kniehl:1998fn} for the respective definitions) are identical at the precision of our computation.
Close to the resonance we can expand $\Sigma_B(E)$ around the complex pole position $E_H+i\Gamma_H/2$ of the hybrid propagator in Eq.~\eqref{HybridProp}. 
In this paper we will consider ultrasoft self-energy contributions $\Sigma^{us}_{B}$ up to two-loop level and NLO in the multipole expansion.
It is therefore sufficient to approximate $\Sigma_B(E_H+i\Gamma_H/2)$ by $\Sigma_B(E^0)$ and set $E_0=V_o^B |_{{\cal O}(\al^2)}$ for $D=4$ and $E_0=V_o^B |_{{\cal O}(\al)}$ for $D=3$. Note that in the latter case ${\cal O}(\al^2)$ terms would produce subleading ${\cal O}(r^3)$ corrections to the energy and decay width.

We now choose to shift the energy origin by adding $-2m_B$, i.e.  $E \to E -2m_B$. The term $2m_B=2m + 2\delta m_{(soft)}$  will allow to absorb ultraviolet soft divergences (into $\delta m_{(soft)}$) that may appear in the computation. $m$ has a clear physical meaning in the case of heavy quarkonium: it represents the heavy quark (pole) mass.
At NLO in the multipole expansion we thus have
\begin{align}
\frac{1}{E-2m_B-V_o^B-\Sigma_B(E-2m_B)}=\frac{Z_H}{E-E_H+i\Gamma_H/2}+{\cal O}( (E-E_H+i\Gamma_H/2)^0)\,,
\end{align}
where $Z_H$ is the normalization of the hybrid propagator and
\begin{align}
E_H-i\Gamma_H/2=m_B+V_o^B+\Sigma_B(E_0)
\label{HybridPole}
\end{align}
is the complex position of the hybrid propagator pole. It can be understood to be composed of contributions from different energy regions. After factorization we can express it in the following way (either bare or renormalized): 
\begin{align}
E_H(r)-i\Gamma_H/2
&=
2m_B+V_{o,B}+\delta E^{us}_{o,B}+\delta E^{np}_{H,B}
\nn
\\
&=
2m_{\MS}(\nu_s)+V_{o,\MS}(r;\nu_s,\nu)+\delta E^{us}_{o,\MS}(\nu,\nu_{np})+\delta E^{np}_{H,\MS}(\nu_{np})
\,.
\label{Estot}
\end{align}
The subscript $\MS$ of $m_{\MS}(\nu_s)$ denotes the scheme for possible hard infrared divergence subtractions (canceling the ultraviolet soft divergences), e.g. in three dimensions, and should not be confused with the ultraviolet (hard) renormalization scheme for the mass. In four dimensions we have $m_{\MS}(\nu_s)=m_{\rm OS}$, where $m_{OS}$ is the usual on-shell mass of the heavy quarks. Similarly $V_{o,\MS}(r;\nu_s,\nu) \rightarrow V_{o,\MS}(r;\nu)$ in 
four dimensions, where $V_o$ is the octet potential and encodes the 
effects due to soft degrees of freedom. $\delta E^{us}_o$ and $\delta E_H^{np}$ encode the effects due to the ultrasoft and nonperturbative degrees of freedom respectively. % with ${\cal O}(r^2)$ precision.
Each term depends on the respective factorization/renormalization scales: $\nu_s$ separates the hard ($m$) from the soft scale ($1/r$), 
$\nu$ separates the soft from the ultrasoft scale ($\Delta V$) 
and $\nu_{np}$ separates the ultrasoft from the nonperturbative scale ($\lQ$). For real factorization scales $\nu_s$, $\nu$, $\nu_{np}$, the decay width 
$\Gamma_H$ is contained in the terms $\delta E^{us}_{o}$ and $\delta E_H^{np}$. We give more details in the next sections. 

\section{Results for $D=4$}
\label{secD4}

In four dimensions the renormalized coupling constant $\al$ satisfies a non-trivial RG equation.
We will work here in the $\MS$ renormalization scheme, which is related to the MS scheme by a redefinition of the renormalization scale: $\nu_{(\rm MS)} \rightarrow \nu_{(\MS)} \, c^{-1}_{\MS}$, where $c_{\MS}=e^{\frac{1}{2}(\ln(4\pi)-\gamma_E)}$. The counterterm coefficient  
\begin{align}
Z^{(1)}_{\al}=\frac{\al}{4\pi}\beta_0+ \frac12 \frac{\al^2}{(4\pi)^2} \beta_1 + \ldots 
\end{align}
is understood in the MS scheme, but the form of the anomalous dimension of $\al$, 
\begin{align}
%\nu \frac{\partial}{\partial \nu} \al = 
\al\beta(\al)=-2\al\left(\beta_0\frac{\al}{4\pi}+\beta_1\frac{\al^2}{(4\pi)^2} + \ldots \right) 
\end{align}
is the same in both schemes, MS and $\msb$. The constants
$\beta_0 = \frac{11}{3}C_A - \frac43 T_F n_f$ and
$\beta_1 = \frac{34}{3}C_A^2 - 4 C_F T_F n_f - \frac{20}{3}C_A T_F n_f$ are the
standard $\msb$ (MS) one- and two-loop coefficients of the QCD beta
function.

We also have to consider possible soft and ultrasoft corrections to $V_A$. They were obtained in Ref.~\cite{RG} with LL accuracy, in Ref.~\cite{Brambilla:2006wp} with NLO accuracy and in Ref. \cite{Brambilla:2009bi} with NLL accuracy.
The outcome is
\begin{align}
\nu\frac{d}{d\nu}V_A=0+{\cal O}(\al^3)
\end{align}
for the anomalous dimension and $V_A=1+{\cal O}(\al^2)$ for the initial matching condition. 
We conclude that for the precision of our calculation we can use $V_A=1$.

The counterterms of the octet potential, which subtract the ultraviolet divergences from the result of our ultrasoft two-loop computation in Eq.~\eqref{USbare2loop} in four dimensions, read
\begin{align}
Z^{(1)}_{V_o}&=-
r^2\Delta V^3 \frac{1}{2N_c}V_A^2\left[
\frac{\al}{3\pi}
+\frac{\alpha ^2 \left[ C_A (47 -12 \pi^2)  - 10 T_F n_f \right]}{108 \pi ^2}
\right]
,
\label{ZVo4D1}\\
Z^{(2)}_{V_o}&=-r^2\Delta V^3  \frac{1}{2N_c}V_A^2\,\frac{2}{3}\, \beta_0 \,\frac{\al^2}{(4 \pi)^2}
\,.
\label{ZVo4D2}
\end{align}
The latter expression comes from the $1/\epsilon_4^2$ pole of the two-loop result and from the $\al^2/\epsilon_4$ divergence of $\al_B$ in the divergent term of the one-loop self-energy.

From Eqs.~\eqref{ZVo4D1} and~\eqref{ZVo4D2} we can derive the RG equation of the octet potential at two-loop order\footnote{In our counting this translates to N$^3$LL order, because $\Delta V$ comes at least with one power of $\al$.}. We find
\begin{align}
\nu\frac{d}{d\nu} V_{o,\MS}
=r^2\Delta V^3 \frac{1}{2N_c}V_A^2
\left[
\frac{2\al}{3\pi}
+
\frac{\alpha ^2 \left[ C_A (47 -12 \pi^2)  - 10 T_F n_f \right]}{27 \pi ^2}
+
{\cal O}(\al^3)
\right].
\label{VoRGE4D}
\end{align}
This result holds in any momentum-independent renormalization scheme that is related to the MS scheme by a ($D$ independent) redefinition of the renormalization scale $\nu$. 
Solving the RG equation we can write the RG improved static octet potential as
\begin{align}
\label{VRGIoMS}
V_{o,\MS}(r;\nu)=V^{i.c.}_{o,\MS}(r;\nu_i)+\delta V_{o,\MS}^{RG}(r;\nu_i,\nu)
\,,
\end{align}
where
\begin{align}
&
\delta V_{o,\MS}^{RG}(r;\nu_i,\nu)
=-\frac{1}{2N_c}
V_A^2
\, r^2 \, (\Delta V)^3 \, \frac{2\pi}{\beta_0}
\\
\nn
&
\times
\left\{
\frac{2}{3\pi}\ln\frac{\al(\nu)}{\al(\nu_i)}
-(\al(\nu)-\al(\nu_i))
\left(
\frac{8}{3(4\pi)^2}\frac{\beta_1}{\beta_0} - \frac{ C_A (47 -12 \pi^2)  - 10 T_F n_f }{27 \pi ^2}
\right)
\right\}
\label{Running4D}
\end{align}
describes the ultrasoft RG evolution of  $V_o$ and the initial matching condition at the (soft) scale $\nu_i$ is given by
\begin{align}
V^{i.c.}_{o,\MS}(r;\nu_i)
&=
 \frac{1}{2N_c}
\frac{\,\alpha(\nu_i)}{r}\,
\sum_{n=0}^{3}\left(\frac{\alpha(\nu_i)}{4\pi}\right)^n a^{(o)}_n(r;\nu_i)
\end{align}
with coefficients ($a_0^{(o)}(r;\nu_i)=1$)
\begin{align}
a_1^{(o)}(r;\nu_i)
&=
a_1+2\beta_0\,\ln\left(\nu_i e^{\gamma_E} r\right)
\,,
\nonumber\\
a_2^{(o)}(r;\nu_i)
&=
a_2^{(o)} + \frac{\pi^2}{3}\beta_0^{\,2}
+\left(\,4a_1\beta_0+2\beta_1 \right)\,\ln\left(\nu_i e^{\gamma_E} r\right)\,
+4\beta_0^{\,2}\,\ln^2\left(\nu_i e^{\gamma_E} r\right)\,
\,,
\nonumber \\
a_3^{(o)}(r;\nu_i)
&=
a_3^{(o)}+ a_1\beta_0^{\,2} \pi^2+\frac{5\pi^2}{6}\beta_0\beta_1 +16\zeta_3\beta_0^{\,3}
\nonumber \\
&+\bigg(2\pi^2\beta_0^{\,3} + 6a^{(o)}_2\beta_0+4a_1\beta_1+2\beta_2+\frac{16}{3}C_A^{\,3}\pi^2\bigg)\,
  \ln\left(\nu_i e^{\gamma_E} r\right)\,
\nonumber \\
&+\bigg(12a_1\beta_0^{\,2}+10\beta_0\beta_1\bigg)\,
  \ln^2\left(\nu_i e^{\gamma_E} r\right)\,
+8\beta_0^{\,3}  \ln^3\left(\nu_i e^{\gamma_E} r\right)\,.
\label{eq:Vr}
\end{align}
Explicit expressions for $a_1$ and $a_2^{(o)}$ can be found in the literature~\cite{FSP,Schroder:1998vy}. The constant $a_3^{(o)}$ is still unknown and represents the only missing piece in the N$^3$LL result for the Wilson coefficient $V_o$, Eq.~\eqref{VRGIoMS}. Note that this expression is real for positive renormalization and matching scales $\nu$ and $\nu_i$.

We can now determine the complex pole of the hybrid (octet) propagator in four dimensions. 
Up to nonperturbative effects, which we neglect here, it is given by
\begin{align}
\lefteqn{ E_H(r) - i\frac{\Gamma_H}{2} \simeq E_o(r) - i\frac{\Gamma_o}{2} \simeq }& \nn\\
&\simeq V^{i.c.}_{o,\MS}(r;1/r)+\delta V_{o,\MS}^{RG}(r;1/r,\Delta V)+\delta E^{us}_{o,\MS}(\Delta V)\,.
\label{Pole4D}
\end{align}
Therefore, besides the octet potential, we also need the ultrasoft correction in the $\MS$ scheme, which we obtain from our bare 
two-loop result in Eq.~\eqref{USbare2loop} after subtraction of the divergences. It reads
\begin{align}
&
\delta E_{o,\MS}^{us} (\nu)
=
\frac{1}{2N_c} r^2(-\Delta V)^3 V_A^2
\bigg[
-\frac{\al }{9 \pi } \left(6 \ln \! \left[\frac{-\Delta V}{\nu } \right]+6\ln 2-5\right)
\label{Eus4Dfinite} \\
&
\nn  
+\frac{\al^2 }{108 \pi ^2} 
\bigg( 18 \beta_0 \ln ^2 \! \left[ \frac{-\Delta V}{\nu} \right]-6 (N_c \left(13-8 \pi ^2\right)-2 \beta_0 (-5+3\ln 2)) \ln \! \left[\frac{-\Delta V}{\nu} \right]
   \\
&   
-2 N_c \left(-84+39 \ln 2-8 \pi ^2 (-2+3\ln 2)+72 \zeta (3)\right)+\beta_0 \left(67+3 \pi ^2-60 \ln 2+18\ln^2 2\right) \bigg)
   \bigg] \nn
   \,.
\end{align} 
Note that this object is complex for positive $\nu$. Since the octet potential is real, the imaginary part of $\delta E_{o,\MS}^{us} (\nu)$ directly gives the decay width.
Choosing $\nu =  \Delta V$ in Eq.~\eqref{Eus4Dfinite} we find for the decay width of the hybrid (octet) system
\begin{align}
\Gamma_H =\frac{4}{3}\al(\Delta V)\frac{1}{2N_c}r^2\Delta V^3
\left[1+\frac{\al(\Delta V)}{12\pi}\left(N_c \left(13-8 \pi ^2\right)-2 \beta_0 (3\ln 2 -5)\right)+ \ord( \al^2)\right] + \ord(r^4)
\,,
\end{align}
where we have to use the NLO expression for the ultrasoft scale $\Delta V$:
\begin{align}
\label{DeltaV4D}
\Delta V(r;\nu_i)
&=\frac{C_A}{2}
\frac{\,\alpha(\nu_i)}{r}\,
\sum_{n=0}^{1}\left(\frac{\alpha(\nu_i)}{4\pi}\right)^n a^{(o)}_n(r;\nu_i)
\,,
\end{align}
which is scheme and factorization scale independent at this order.

For $E_H$ we need the real part of $\delta E_{o,\MS}^{us} (\nu)$:
\be
E_H(r) = V^{i.c.}_{o,\MS}(r;1/r)+\delta V_{o,\MS}^{RG}(r;1/r,\Delta V)+{\rm Re}\,\delta E^{us}_{o,\MS}(\Delta V)\,.
\ee
Replacing the first two terms in this equation by~Eq.\eqref{VRGIoMS} and using the leading term of Eq.~\eqref{Eus4Dfinite} in the last term, we reach N$^3$LL accuracy. Eq.~\eqref{Eus4Dfinite} also provides the subleading ultrasoft correction to the hybrid energy relevant at N$^4$LL order.

We will not attempt to compare our results with 4D lattice simulations. This would require the incorporation of nonperturbative effects and the proper treatment of renormalons,
which goes beyond the scope of this paper. 

\section{Results for $D=3$}
\label{sec3D}

The derivation of the static octet potential in three dimension is quite analogous to the color singlet case, which has been discussed in detail in Ref.~\cite{Pineda:2010mb}. We therefore focus on the novel aspects of the analysis for the octet (hybrid) system in this section and refer to Ref.~\cite{Pineda:2010mb} for a more careful account on the universal issues related to the 3D static potential. 

As argued in Ref.~\cite{Pineda:2010mb}, the coefficients $V_{A/B}$ are not renormalized at $\ord(r^0)$, i.e.
\begin{align}
\label{ZVAB}
Z_{A/B}=1+ \ord(r)
\,.
\end{align}
The reason is that the potentials and $\al$ have to appear perturbatively (with positive powers) in the counterterms, otherwise the renormalizability of the theory at leading order of the multipole expansion would be spoiled. 
Moreover, we can set $V_A=V_B=1$ (just like in 4D), as ${\cal O}(\al)$ soft corrections would be multiplied by factors of $r$ and would move us away from 
the precision of $V_o$ aimed for at this paper. Actually, from inspection of the possible diagrams that will contribute 
at the soft scale, we know that $V_{A/B}=1+{\cal O}(\al^2)$~\cite{Brambilla:2006wp}.

From the bare soft computation we can completely fix $\delta V_o$ through ${\cal O}(r^2)$. 
We obtain (for the ultrasoft counterterms)
\begin{align}
\label{Z1Vo}
Z^{(1)}_{V_o}
&=
\frac{C_A}{2}\, \al +r^2 \Delta V^2\al \Big(C_F-\frac{C_A}{2}\Big) \frac{1}{4}
+r^2 \Delta V \, C_A\al^2 \Big(C_F-\frac{C_A}{2}\Big) \frac{1}{2}
\\
\nn
&
-r^2\al^3 \Big(C_F-\frac{C_A}{2}\Big) 
%  \left(
\\
\nn
&
\times\frac{\left(13 \pi ^2-2304\right) C_A^2+8 \left(19\pi^2+144\right) 
C_A T_Fn_f-48 T_Fn_f \left(4 \left(\pi ^2-10\right) C_F+\pi ^2T_Fn_f\right)}{2304}\,,
%   \right)
\end{align}
\begin{align}
\label{Z2Vo}
Z^{(2)}_{V_o}
=
r^2 \Delta V \al^2 \Big(C_F-\frac{C_A}{2}\Big) C_A\frac{1}{8}
+r^2\al^3 \Big(C_F-\frac{C_A}{2}\Big) C_A^2\frac{1}{12}\,,
\end{align}
\begin{align}
\label{Z3Vo}
Z^{(3)}_{V_o}
=r^2\al^3 \Big(C_F-\frac{C_A}{2}\Big) C_A^2\frac{1}{48}\,,
\end{align}
\begin{align}
\label{ZnVo}
Z^{(n)}_{V_o}=0 \quad \forall \quad n>3\,.
\end{align}
The soft calculation is organized in powers of $\al\, r$. The tree level computation gives the first term in 
Eq.~\eqref{Z1Vo} (once the ultraviolet divergences of the soft one-loop heavy quark self-energies have been subtracted). The soft one-loop contribution to the octet potential is infrared safe. The two-loop 
computation leads to the remaining terms. These results are exact at ${\cal O}(r^2)$. 

The fact that one can renormalize the 
potential with a finite number of terms at a given order in the multipole expansion, i.e. $Z^{(n)}_{V_o}=0$ for $n>3$ at ${\cal O}(r^2)$, 
reflects the super-renormalizability of the theory in three dimensions. 

\begin{figure}
\includegraphics[width=0.25 \textwidth]{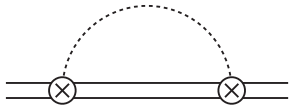}
%\makebox[0.0cm]{\phantom b}
%\epsfxsize=5truecm \epsfbox{usoctet3D.eps}
\caption{\it One-loop contribution to the octet propagator at ${\cal O}(r^0)$. The dotted line 
represents the $A^0$ field.}
\label{usoctet3D}
\end{figure}

The infrared divergences of the bare potential have to cancel the ultraviolet ones of the ultrasoft computation. We can then use 
the explicit one- and two-loop ultrasoft contributions to partially check the structure of the octet potential counterterm in Eqs.~\eqref{Z1Vo}-\eqref{Z3Vo}. 

At leading order in the multipole expansion the octet field has 
a residual interaction with the ultrasoft gluon field. 
The octet potential receives an ultraviolet divergent correction from 
the one-loop self-energy diagram shown in Fig. \ref{usoctet3D}, which yields the first term in Eq.~\eqref{Z1Vo}. 
This is the only term at $\ord(r^0)$, since higher loop diagrams cannot contribute because the coupling and the potentials have to appear perturbatively (with positive powers) in the Z's. Since $\al$ has dimensions of mass, the potentials would appear with negative powers in multi-loop diagrams at $\ord(r^0)$. This 
is not allowed by renormalizability. By the very same reason the octet field does not require renormalization at $\ord(r^0)$:
\begin{align}
Z_o=1 + \ord(r^2).
\end{align}

We now consider higher orders in the multipole expansion. 
 At two soft loops in dimensional regularization infrared poles up to $\ord(1/\epsilon_3^3)$ appear.
The ultrasoft computation in pNRQCD yields the following results for the counterterms: 

1) The second term in Eq.~\eqref{Z1Vo} comes from the $1/\epsilon_3$ divergence of the ultrasoft one-loop correction in Eq.~\eqref{USbare1loop}.
It is scheme independent and fixes, together with Eqs.~\eqref{Z1} and~\eqref{Z2}, the first term of Eq.~\eqref{Z2Vo} and 
Eq.~\eqref{Z3Vo}. It would also be possible to compute these $1/\epsilon_3^2$ and $1/\epsilon_3^3$
divergences directly. The $1/\epsilon_3^2$ term is confirmed by the ultrasoft two loop result in Eq.~\eqref{USbare2loop}. The $1/\epsilon_3^3$ term would however require an ultrasoft three-loop calculation, which has not been performed yet.

2) The third term in Eq.~\eqref{Z1Vo} follows from the remaining $1/\epsilon_3$ divergence in the ultrasoft two-loop correction, Eq.~\eqref{USbare2loop}, once all subdivergences (associated with the $\ord(r^0)$ octet potential) have been subtracted. 
This result combined with Eq.~\eqref{Z2} then fixes the 
second term of Eq.~\eqref{Z2Vo}.

In summary: the explicit ultrasoft computation allows to check the three first terms of Eq.~\eqref{Z1Vo} and the first term of Eq.~\eqref{Z2Vo}. 
We find perfect agreement. The use of Eqs.~\eqref{Z1} and~\eqref{Z2} allows us to completely check Eq.~\eqref{Z2Vo} and~\eqref{Z3Vo}. 
We also find perfect agreement. Note that this can be understood as 
a non-trivial cross-check of two independent determinations of these terms.

The only remaining term is the last one in Eq.~\eqref{Z1Vo}. 
This term would be canceled by the sum of the ultrasoft ultraviolet
divergences from three-loop octet self-energy diagrams with and without singlet-octet transition. The latter diagrams are proportional to $V_B^2$ or $V_C$
and scaleless (for onshell external fields), as they are insensitive to the ultrasoft scale $\Delta V$. Therefore, the associated RG evolution will run down to the non-perturbative scale $\al$. This is not so for the ultrasoft diagrams with singlet-octet transition, which get infrared regulated by the
ultrasoft scale $\Delta V$ correspondingly producing logarithms of $\Delta V$. Without an explicit ultrasoft three-loop calculation we are not able to distinguish among the divergences from the two types of diagrams nor check the last term of Eq.~\eqref{Z1Vo}, which has been deduced from the soft result. Therefore, we turn the problem around and use the latter to fix the ultraviolet divergences of the ultrasoft three-loop computation for the next sections.

\subsection{pNRQCD RG}

We can now deduce the RG equations of $V_o$. Again the discussion is similar to the one of 
Ref. \cite{Pineda:2010mb} to which we refer for extra details.

At leading order in the multipole expansion the RG equation of the octet potential reads 
\begin{align}
\nu\frac{d}{d\nu}V_o=-C_A\al\,.
\end{align}
The running of $\Delta V$ at this order reads (including the tree level matching condition) 
\begin{align}
\label{DeltaVoRG}
\Delta V_X(r;\nu)=-\al C_A\ln(r\nu d_X)+{\cal O}(\epsilon)
\,,
\end{align}
where $X$ stands for the factorization scheme (e.g. MS or $\MS$):
\begin{align}
d_{\rm MS}=e^{\gamma_E/2}\sqrt{\pi} \simeq 2.36546,
\qquad 
d_{\rm \MS}=d_{\rm MS}\, c_{\MS}^{-1}=e^{\gamma_E}/2 \simeq 0.890536,
\end{align}
with $c_{\MS}= e^{1/2(\ln (4\pi) - \gamma_E )}$.

Now we consider the subleading contributions to $V_o$.
The {\it complete} anomalous dimension of the static octet potential through ${\cal O}(r^2)$ takes the form
\begin{align}
\nu\frac{d}{d\nu}V_o=B(V_o)=-C_A\al+\sum_{n=0}^2C_F\al 
V_A^2r^2\al B_n(C_A\al)^n(\Delta V)^{2-n},
\end{align}
and is explicitly given by
\begin{align}
\label{VoRGeq}
\nu\frac{d}{d\nu}V_o &=-C_A\al-
 r^2 \Delta V_X^2\al \Big(C_F-\frac{C_A}{2}\Big) V_A^2\frac{1}{2}
-2r^2 \Delta V_X \, C_A\al^2 \Big(C_F-\frac{C_A}{2}\Big) V_A^2
\\
\nn
&
+r^2\al^3 \Big(C_F-\frac{C_A}{2}\Big) V_A^2
%  \left(
\\
\nn
&
\times
\frac{\left(13 \pi ^2-2304\right) C_A^2+8 \left(19\pi^2+144\right) 
C_A T_Fn_f-48 T_Fn_f \left(4 \left(\pi ^2-10\right) C_F+\pi ^2T_Fn_f\right)}{384}\,.
%   \right)
\end{align}
The form of the anomalous dimension in Eq.~\eqref{VoRGeq} is invariant under scheme transformations that amount to a redefinition of the renormalization scale: $\nu \rightarrow \nu\, c_X^{-1}$, where $c_X$ is a $\epsilon_3$-independent constant ($c_{\rm MS}=1$). We indicate renormalized quantities in this class of momentum independent renormalization schemes by an index ``$X$'' (X=MS, $\msb$, etc.) as e.g. $\Delta V_X$ in the above equations.

By solving the RG equations in the MS scheme we obtain the static octet potential with ${\cal O}(\al^3 r^2)$ precision:
\begin{align}
\label{VoRG}
V_{o,{\rm MS}}(r;\nu_s,\nu)=V^{i.c.}_{o,{\rm MS}}(r;\nu_s, \nu_i)+\delta V^{RG}_{o,{\rm MS}}(r;\nu_i,\nu)\,,
\end{align}
where
\begin{align}
\delta V^{RG}_{o,{\rm MS}}(r;\frac1{r},\nu)=&-C_A\al\ln(\nu r)+\Big(C_F-\frac{C_A}{2}\Big) r^2 \alpha ^3  \nonumber\\
&\times
 \Bigg\{-\frac{1}{6} C_A^2 \ln^3(r \nu ) 
%\nonumber\\ &
-\frac{1}{4} C_A^2 \big[-4+\gamma_E +\ln \pi \big] \ln^2(r \nu ) \nonumber\\
&
+\bigg[ C_A^2 \Big(\frac{13\pi ^2}{384}+\frac{1}{8} \big(-\gamma_E ^2 + 8 \gamma_E - 48 + (8 - 2 \gamma_E)\ln \pi - \ln^2 \pi   \big) \Big) \nonumber\\
&
+n_f T_F \Big(C_A \Big(3+\frac{19 \pi ^2}{48}\Big) +C_F \Big(5-\frac{\pi ^2}{2}\Big)\Big) - (n_f T_F)^2 \frac{ \pi ^2}{8}\bigg] \ln (r \nu)
\Bigg\}
\label{Vorunning}
\end{align}
is the running and 
\begin{eqnarray}
  \lefteqn{ V_{o,{\rm MS}}^{i.c.}(r;\nu_s, \frac1{r}) = 
C_F \alpha \ln (r^2 \nu^2_s\pi e^{\gamma_E} )-\frac{C_A}{2} \alpha \ln (\pi e^{\gamma_E} )  
+  \frac{\pi}{4} \Big(C_F-\frac{C_A}{2}\Big) (7 C_A-4 n_f T_F) \alpha^2 r
   } \nonumber\\
&&+\Big(C_F-\frac{C_A}{2}\Big) r^2 \alpha ^3  
 \Bigg\{C_A^2 \bigg[\frac{\pi ^2}{2304} (39 \gamma_E-715 +564 \ln 2+39 \ln \pi ) + \frac{25}{3}-\frac{31}{24} \zeta (3) \nonumber\\
&&
 \qquad \qquad-\frac{1}{48} (\gamma_E +\ln \pi ) \left(144-12 \gamma_E +\gamma_E ^2+\ln ^2\pi -12 \ln \pi +2 \gamma_E  \ln \pi \right) \bigg] \nonumber\\
&&
\qquad
 + n_fT_F C_F \Big[\frac{\pi^2}{24}  (15-6 \gamma_E -8 \ln 2-6 \ln \pi )+\frac{1}{2} (5 \gamma_E-8 +5 \ln \pi )\Big] \nonumber\\
&&
\qquad
+ n_fT_F C_A \Big[\frac{\pi^2}{288} (57 \gamma_E -97 +12 \ln 2+57 \ln \pi )+\frac{1}{6} (9 \gamma_E -22 +9 \ln \pi +10 \zeta (3))\Big] \nonumber\\
&&
\qquad
+(n_f T_F)^2 \Big[\frac{\pi^2}{48}(7-3 \gamma_E -\ln (16 \pi ^3) ) \Big] \Bigg\}
\label{Vo1r}
\end{eqnarray}
is the initial matching condition. Note that the tree level and one-loop matching conditions have been included. The tree-level result depends on the factorization scale $\nu_s \gg 1/r$, which separates the hard and the soft regime. The dependence on this factorization scale cancels the infrared scale dependence of the mass, cf. Eq.~\eqref{Estot}.

We stress that we have obtained the {\it exact} contributions to the static potential through $O(r^2)$. There is nothing left. Moreover, by setting $\nu \sim \Delta V$ large logarithms are resumed up to the scale $\Delta V$, i.e. we have determined all logarithms of $r \Delta V$ at $O(\al^3 r^2)$. 

The above results have been presented in the MS scheme. We can easily transform them to another momentum independent scheme by redefining the renormalization scale. In particular, if we write the expression in terms of $\Delta V$, most of the scheme dependence gets encapsulated in $\Delta V$. For instance,
Eq.~\eqref{Vorunning} can be reexpressed as
\begin{align}
\nn
&
\delta V^{RG}_{o,{\rm MS}}(r;\frac1{r},\nu)
=
\Delta V_X(r;\nu)-\Delta V_X(r;1/r)
+ \frac{1}{6} \Big(\frac{C_F}{C_A}-\frac{1}{2}\Big)  r^2 \al  V_A^2\left(\Delta V^3_X(r;\nu)-\Delta V^3_X(r;1/r)\right)
\\
\nn
&
+r^2 \, C_A\al^2 \Big(\frac{C_F}{C_A}-\frac{1}{2}\Big) V_A^2\left(\Delta V^2_X(r;\nu)-\Delta V^2_X(r;1/r)\right)
\\
&
\nn
-r^2\al^2 \Big(\frac{C_F}{C_A}-\frac{1}{2}\Big) V_A^2
 \\
\nn
&
\times
 \left(
\frac{\left(13 \pi ^2-2304\right) C_A^2+8 \left(19\pi^2+144\right) 
C_A T_Fn_f-48 T_Fn_f \left(4 \left(\pi ^2-10\right) C_F+\pi ^2T_Fn_f\right)}{384}
   \right)
   \\
   &
   \qquad
   \times \left(\Delta V_X(r;\nu)-\Delta V_X(r;1/r)\right)
\,.
\end{align}

\subsection{Ultrasoft contributions ($\delta E^{us}_o$)}

The general structure of  the ultrasoft contribution $\delta E^{us}_o$ at ${\cal O}(r^2)$ is
\begin{align}
\delta E_o^{us}(\nu,\nu)
=
-\frac{1}{2N_c}\al\, V_A^2 \, r^2 \Delta V^2 \,\sum_{n=0}^{\infty}\left(\frac{C_A\al}{\Delta V}\right)^n\sum_{s=0}^{n+1}c_{n,s}\ln^s\Big[\frac{\Delta V}{\nu}\Big]
\,,
\label{USstruc}
\end{align}
where we have set $\nu_{np}=\nu$. 
The dependence on $\nu_{np}$ first appears at three loops in this ${\cal O}(r^2)$ contribution to  $\delta E^{us}_o$. 
At present, concrete results for the ultrasoft corrections are available at one, $\ord(g^2)$, and two loop level, $\ord(g^4)$, given in Eqs.~\eqref{USbare1loop} and \eqref{USbare2loop}. After MS subtraction they read 
\begin{align}
\delta E^{us}_{o,\rm MS}(\text{1-loop})
=
\frac1{8N_c} \al\, V_A^2 r^2\, \Delta V_{\rm MS}^{2}
\bigg(
1+\gamma_E-\ln(4\pi)+2\ln \Big[ \frac{-\Delta V_{\rm MS}}{\nu} \Big]
\bigg),
\label{deltaEus3D1loop}
\end{align}
\begin{align}
\nn
\delta E^{us}_{o,\rm MS}(\text{2-loop})
&=
\frac{1}{4 N_c} \al^2\, V_A^2\, r^2 \, \Delta V_{\rm MS} \bigg(-C_A \ln^2 \Big[ \frac{-\Delta V_{\rm MS}}{\nu} \Big]
\\
\nn
&
+ 
C_A (2-\gamma_E +\ln (4 \pi )) \ln \Big[ \frac{-\Delta V_{\rm MS}}{\nu} \Big]\\
& - \frac{1}{2} C_A \big(\pi ^2 -7 +\frac{1}{2} ( 2-\gamma_E +\ln (4 \pi) )^2 \big) - 2\, n_f T_F \bigg).
\label{deltaEus3D2loop}
\end{align}
Note that the these expressions have imaginary parts. 
Setting $\nu \sim \Delta V_X(\nu)$ resums large logarithms into the potential $V_o$ and minimizes them in the ultrasoft contribution $\delta E^{us}_{o}$.
It also simplifies the determination of the decay width of the octet (hybrid) system:
\begin{align}
\Gamma_H =
\frac{1}{2N_c} \al\, V_A^2\, r^2 \Delta V_{\rm MS}^{2}(\bar \nu_{us})\,\pi
+\frac{1}{2 N_c} \al^2\, V_A^2\, r^2 \Delta V_{\rm MS}(\bar \nu_{us})\, C_A \pi(2 \!-\! \gamma_E \!+\! \ln (4 \pi )) + \ord(\al^3 r^2 \Delta V^0)
\label{GammaH3D}
\end{align}
where
\begin{align}
\bar \nu_{us} \equiv \Delta V_X(\bar \nu_{us})=C_A\al\, W(1/(C_A\al\, d_X r))\,
\end{align}
and $W(z)$ is the Lambert function (here: $X=\rm MS$).  

The precision of Eq.~\eqref{GammaH3D} is limited by the existence of some unknown single logarithms that appear in the three-loop ultrasoft self-energy (see next subsection). The imaginary parts of those logarithms produce terms like $\delta \Gamma \sim \al^3r^2$.

\subsection{Subleading ultrasoft and nonperturbative effects}

Even without an explicit computation we can obtain some information on the ultrasoft three-loop terms. At three loops we start to have contributions from diagrams with no singlet-octet vertices. They are proportional to $V_B^2$ or $V_C$ and vanish in dimensional regularization (for onshell external fields), as they are insensitive to the scale $\Delta V$. Therefore, any ultraviolet divergence proportional to $V_B^2$ or $V_C$ should be canceled by an infrared one, which signals a sensitivity to the nonperturbative scale $\al$. We quantify this 
 statement with the following equation
\begin{align}
\nu\frac{d}{d\nu}\delta E^{us}_o(\nu,\nu)=-B(V)-\nu\frac{d}{d\nu}\delta E^{np}_H\,,
\end{align}
where the RG structure of the nonperturbative term is the following
\begin{align}
\nu\frac{d}{d\nu}\delta E^{np}_H=C_A\al +B\,r^2\al^3(1+{\cal O}(\al/\Delta V))\,.
\end{align}
Solving this equation we obtain
\begin{align}
\delta E^{np}_H (\nu_{np}) =C_A\al\,(\ln \Big(\frac{\nu_{np}}{\al} \Big)+c_H)+B\al^3r^2\ln \Big(\frac{\nu_{np}}{\al} \Big)+{\cal O}(\al^3r^2)\,,
\label{EnpH3D}
\end{align}
where $c_H$ is a nonperturbative constant that depends on the specific hybrid (gluelump) we consider. Actually, the ${\cal O}(r^0)$ term is nothing but the gluelump mass
\be
\Lambda_H(\nu_{np})=C_A\al\,(\ln \Big(\frac{\nu_{np}}{\al} \Big)+c_H)
\,.
\ee
It can be related to a gauge invariant correlator as
\begin{align}
\Lambda_H = \lim_{T \to \infty} \frac{i}{T} \ln\, \langle H^a(T/2) \, \phi^{\rm adj}_{ab}(T/2,-T/2) \,H^b(-T/2) \rangle\,,
\end{align}
where $H^a$ is the field operator associated with the gluonic component of the hybrid and $\phi^{\rm adj}_{ab}(T/2,-T/2)$ is a Wilson line in the adjoint representation, see Ref.~\cite{Brambilla:1999xf}.

Note on the other hand that $B$ is independent of the hybrid type and can be obtained from perturbation theory. It is however unknown at present and constrains, besides the nonperturbative gluelump mass $\Lambda_H $, the precision of our result.

At this point we could also analyse subleading ultrasoft effects in the $\al/\Delta V$ expansion along the lines of Ref.~\cite{Pineda:2010mb}, as some of them, namely the logarithmic terms proportional to $V_A^2$, are fixed by the RG structure. In view of the dominant uncertainties discussed above, and because there are more unknowns than in the singlet case~\cite{Pineda:2010mb}, we refrain from performing that analysis here.

Combining Eqs.~\eqref{VoRG}, \eqref{deltaEus3D1loop}, \eqref{deltaEus3D2loop} and~\eqref{EnpH3D} we can write down the static hybrid energy $E_H= 2 m_{\rm MS} +  V_{o,{\rm MS}} + \delta E_{o, \rm MS}^{us} + \delta E^{np}_{H, \rm MS}$. 
Expressed as a double expansion in $\al r$ and $1/\ln(r\Delta V)$ it reads with ${\cal O}(\al^2r^2)$ and NNLL accuracy
\begin{align}
E_H =&\; 2m_{\rm MS}(\nu_s)+ C_F \alpha \ln (r^2 \nu^2_s\pi e^{\gamma_E} ) +  \frac{\pi}{4} \Big(C_F-\frac{C_A}{2}\Big) (7 C_A-4 n_f T_F) \alpha^2 r \nn\\
&+\delta V^{RG}_{o,{\rm MS}}(r;\frac1{r},\bar \nu_{us}) +\delta E_{o, \rm MS}^{us}(\bar \nu_{us},\nu_{np}) + \delta E^{np}_{H, \rm MS}(\nu_{np}) + {\cal O}(\al^3 r^2 \ln^0) \nn\\
=&\; 2m_{\rm MS}(\nu_s)+ C_F \alpha \ln (r^2 \nu^2_s\pi e^{\gamma_E} ) -\frac{C_A}{2} \alpha \ln (r^2 
\al^2 \pi e^{\gamma_E} ) + C_A\al\, c_{H,{\rm MS}} \nn\\
 &+ \frac{\pi}{4} \Big(C_F-\frac{C_A}{2}\Big) (7 C_A-4 n_f T_F) \al^2 r
\nn\\
&-\Big(C_F-\frac{C_A}{2}\Big) \alpha ^3\, r^2
\Bigg\{
\frac{1}{6} C_A^2 \ln ^3(r \Delta V_{\rm MS}) + \frac{1}{4} C_A^2 (2 \gamma_E-3 -2\ln 2 ) \ln ^2(r \Delta V_{\rm MS}) \nonumber\\
&\quad +\bigg[C_A^2 \Big(\frac{83 \pi ^2}{384}+\frac{1}{8} \left(4 \gamma_E ^2+4 \ln ^2 2-\gamma_E  (10+\ln 256) +38 + 8 \ln 2 -2 \ln \pi \right)\Big)\nonumber\\
&
\qquad - n_f T_F \Big( C_A \big(2+\frac{19 \pi ^2}{48}\big) + C_F \big(5-\frac{\pi ^2}{2}\big)\Big) + (n_f T_F)^2 \frac{ \pi
   ^2}{8}\bigg] \ln (r \Delta V_{\rm MS}) \Bigg\} \nn\\
&+{\cal O}(\al^3 r^2 \ln(\Delta V_{\rm MS}/\al))
\,. \label{EHNNLL3D}
\end{align}
We have checked that the explicit scheme dependence of $\delta V_{o,{\rm MS}}^{RG}$ and $\delta E^{us}_{o,{\rm MS}}$ and the implicit scheme dependence of Eq.~\eqref{EHNNLL3D} through the logarithms of $\Delta V$ cancel up to $\ord(\al^3 r^2 \ln^0)$. Note also that $c_{H,{\rm MS}}$ is scheme dependent.

The leading uncertainty of Eq.~\eqref{EHNNLL3D} comes from the nonperturbative gluelump mass $\Lambda_H$. The constant $\Lambda_H$ is independent on the distance $r$ and therefore drops out in the force, i.e. the derivative of the potential. In that case the leading uncertainty is due to the coefficient $B$, which is unknown but, unlike $\Lambda_H$, of perturbative origin and could be determined by an ultrasoft three loop calculation. 

Eq.~\eqref{EHNNLL3D} represents one of the main results of this paper. One could try to perform some comparison with the (quenched) lattice data existing in the literature~\cite{Juge:2003ge,Caselle:2004er,Kuti:2005xg,Philipsen:2002az,Brandt:2009tc}. It is not clear though that these simulations reach short enough distances so that our results can be tested quantitatively (on the other hand it should not be too costly to perform a dedicated short distance simulation to test our expression). Moreover, it would be as well very interesting to study the accidental symmetries and degeneracies among different hybrid energies that should appear at short distances along the lines of the studies in four dimensions performed in Refs.~\cite{Brambilla:1999xf,Bali:2003jq}. To this end one could consider energy differences where the perturbative terms (the non-analytic, i.e. logarithmic dependence in $r$) cancels. 
The non-canceled terms should be produced by the nonperturbative terms (note that the scheme dependence of $c_{H/H'}$ cancels in the difference):
\be
E_H-E_{H'}=C_A\al(c_H-c_{H'})+C_{HH'}r^2 + \ord(r^3)
\,.
\ee
All this would require a dedicated analysis, which will be carried out elsewhere. 

\section{Results for $D=2$}
\label{secD2}

It is also interesting to investigate static hybrid systems in two space-time dimensions. In the following we set $n_f =0$ for simplicity.
In exactly two dimensions physical hybrids do not exist, as there are no propagating (physical) gluons. 
It is instructive to see how this finding arises in an explicit calculation. For a typical hybrid, the object to be computed is e.g.
\begin{align}
\langle W_H \rangle = \langle\, \frac1{N_c} { \rm Tr} \;{\cal P}\; {\bf E}(T/2,{\bf R})\cdot {\bf E}(-T/2,{\bf R}')\exp\big(-i g \oint_{\Gamma} A_\mu dx^\mu\big) \,\rangle\,,
\label{octetWilsonloop}
\end{align}
where the contour $\Gamma$ of the integral in Eq.~\eqref{octetWilsonloop} is a rectangle with spatial extension $r$ and temporal extension $T$ in Minkowski space and we have inserted two chromoelectric fields on each space-like end-string of the rectangular Wilson loop. We could have also chosen other gluonic configurations instead, but this would only make the discussion more complicated without changing the physical outcome. 

Eq.~\eqref{octetWilsonloop} is gauge invariant. Choosing axial ($A_1 \equiv 0$) gauge in exactly two dimensions, the only non-vanishing component of the gluon field-strength tensor is $F_{01}=-F_{10}=-\partial_1 A_0$. Hence, ($A_0$) gluons neither interact among themselves, nor propagate in time, since no time derivative acting on the gluon field is left in the Lagrangian. 
Therefore only\footnote{We do not include in this discussion possible interactions of ${\bf E}$ with the gluons in the time-like strings. They would produce terms proportional to the singlet: $\sim e^{iV_sT}$, but not contribute to the hybrid energy.} planar ``ladder'' diagrams (with ``potential'' gluons) contribute to $\langle W_H \rangle$. 
In the large $T$ limit the final result takes the form
\begin{align}
\langle W_H \rangle \sim \langle E^a(T/2) E^a(-T/2) \rangle e^{-iV_o T} 
\label{WilsonD2}
\end{align}
where $V_o(r)=\frac{1}{2N_c} \al \, r$.
Note that this octet potential is the same as obtained from a leading order, tree-level, computation in 2D static QCD. 
Therefore the non-existence of the hybrid does not arise from the fact that the octet potential diverges in two dimensions (generating an infinite hybrid mass) but rather from the fact that 
\begin{align}
 \langle E^a(T/2) E^a(-T/2) \rangle \sim \delta (T)
 \label{ED2}
\end{align}
is local in time. That is because the gluon does not propagate in time. 

For $D > 2$ one expects $\langle W_H \rangle$ to converge towards the previous result in Eqs.~\eqref{WilsonD2}, \eqref{ED2} as $D \rightarrow 2$, provided the limit is smooth. 
The discussion parallels to some extent the one for the singlet potential in Ref.~\cite{Pineda:2010mb}. Like in that case we cannot make definite statements, since the calculation of the Wilson loop becomes intrinsically nonperturbative. The reason is that the proper ultrasoft expansion parameter is now $ \Delta V / g \sim g r$, see Ref.~\cite{Pineda:2010mb}. Thus the nonperturbative ultrasoft ($\Delta V / g$) expansion and the perturbative multipole ($g r$) expansion mix.

All we can say is that the cancelation of the soft and ultrasoft one-loop contributions we observed for the singlet potential~\cite{Pineda:2010mb} also occurs for the hybrid potential, because the computation differs only in the prefactor.
For the chromoelectric correlator we find at leading order in $\al$
\begin{align}
\langle E^a(T/2) E^a(-T/2) \rangle \sim {\cal O}(\epsilon) \,.
\end{align}
This is what we would expect from the decoupling of gluons in two dimensions. Nevertheless at NLO in $\al$ one obtains a finite contribution. Again we can not draw any 
definite conclusion, as this result is obtained in standard perturbation theory, whereas the correct expansion entails inverse powers of the coupling constant $g$.

\section{Conclusions}
\label{conclusion}

We have computed the energy of a static hybrid in the weak coupling limit in four and three space-time dimensions.
Employing finite temparature theory methods we have checked the static potential of a color octet quark-antiquark pair with distance $r$ at two loops for an arbitrary number of dimensions $D$. 
The result represents the soft contribution to the static hybrid energy.
Using the effective theory pNRQCD we have explicitly calculated the missing ultrasoft two-loop contributions. We have also confirmed the respective two-loop result for the color singlet. The cancelation of soft infrared and ultrasoft ultraviolet divergences is a strong cross-check of our results.

For $D=4$ we have determined the static octet potential and the static hybrid energy through $\rm N^3LL$ order, up to the unknown three-loop soft matching condition. 
The result disagrees with an earlier result obtained in Ref.~\cite{Brambilla:2009bi}. 
We have also given the ultrasoft contribution to the static hybrid energy at $\rm N^4LL$ and computed the decay width of the hybrid/octet system at NLL order.

In $D=3$ dimensions our result for the static hybrid energy reaches ${\cal O}(\al^3r^2)$ accuracy in the (soft) multipole expansion. At this order we have determined the complete expression at NNLL order in the ultrasoft $\al/\Delta V$ expansion. The precision of the result is only limited by unknown terms of order ${\cal O}(\al^3r^2\ln(\Delta V/ \al))$ 
and the nonperturbative gluelump mass. The former can in principle be obtained by a perturbative three-loop computation, whereas the latter requires lattice simulations (note, however, that the gluelump mass vanishes in derivatives of the potential, like the force). Besides the energy we have determined the 3D hybrid/octet decay width through ${\cal O}(\al^3r^2)$ in the multipole expansion and NLO in the $\al/\Delta V$ expansion.

We have also studied the two dimensional case, where the exact result is known: hybrids do not exist. This is due to the temporal locality of the gluonic correlators as there are no propagating (physical) gluons in {\it exactly} two space-time dimensions, see Eqs.~\eqref{WilsonD2} and~\eqref{ED2}, rather than due to an infinite octet potential or hybrid mass.
In the $D \rightarrow 2$ limit this result is more complicated to obtain, because already the ultrasoft contribution is intrinsically nonperturbative.
Strong cancelations have to occur among the different contributions from the soft, the ultrasoft and the nonperturbative ($g$) scale. 
We can only report partial and inconclusive results on this limit, particularly because they are based on perturbation theory, which is not reliable for $D \rightarrow 2$. 
%We have found a nice cancellation between the one loop soft and one loop ultrasoft $D$ dimensional contribution to the
%octet static energy in the $D \rightarrow 2$ limit.
%We are not sure of the relevance of this cancellation, as it follows from a perturbative ultrasoft computation. The gluonic 
%correlator vanishes at LO in $\al$ but not beyond.
Nevertheless, it might be worth exploring the $D \rightarrow 2$ limit in more detail, as it could provide nontrivial information about the dependence of the ($D$ dimensional) perturbative results on $(D-2)$.
\bigskip

\acknowledgments{
This work was partially supported by the Spanish 
grants FPA2007-60275 and FPA2010-16963, and by the catalan grant SGR2009-00894.
The Feynman diagrams in this paper have been drawn using {\tt JaxoDraw}~\cite{JaxoDraw}.
}

%\vfill
%\newpage

\appendix
\section{Two-loop pNRQCD self-energy diagrams in Feynman gauge}
\label{diagrams}

\subsection{Color octet}
\label{octdiagrams}

\begin{align}
\lefteqn{\raisebox{-0.5 ex}{\includegraphics[width=0.2 \textwidth]{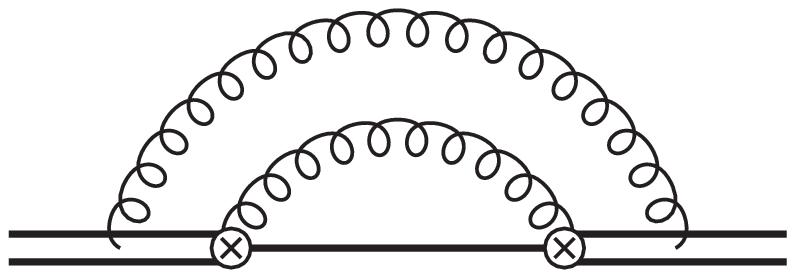} } = C_o^{(d)}\, 
\frac{\Gamma (5-2 d) \Gamma \left(\frac{d}{2}\right)^2}{4 (d-4) (d-3)}  }&&    \\
&=\left\{ \begin{array}{l}
C_o^{(3)}\Big[ \frac{1}{8 \epsilon_3 ^2}+\frac{2 L+1}{4 \epsilon_3 } + L^2 +L+\frac{11 \pi ^2}{48}+1 + \ord(\epsilon_3) \Big]
\\[3 ex]
C_o^{(4)}\Big[-\frac{1}{12 \epsilon_4 ^2}+\frac{-6 L+11-6 \ln 2}{18 \epsilon_4 } -\frac{2 L^2}{3} -\frac{2L (6 \ln 2-11)}{9} -\frac{\pi
   ^2}{8}-\frac{349}{108}-\frac{2 \ln^2 2}{3}+\frac{22 \ln 2}{9}  + \ord(\epsilon_4) \Big]
 \end{array}
\right.
\nn\\[4 ex]
\lefteqn{\raisebox{-3.5 ex}{\includegraphics[width=0.2 \textwidth]{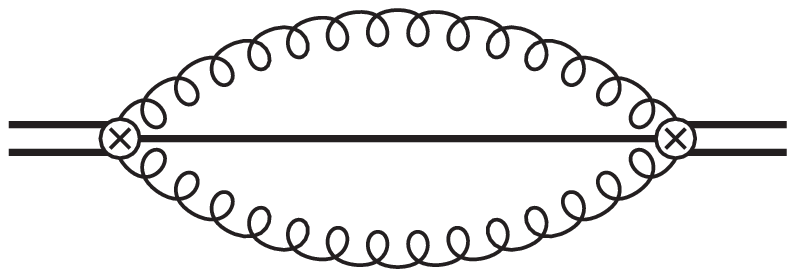} } = C_o^{(d)}\, 
\frac{\Gamma (5-2 d) \Gamma \left(\frac{d}{2}-1\right)^2}{16}  } && \\
&=\left\{ \begin{array}{l}
C_o^{(3)}\Big[-\frac{1}{4 \epsilon_3 } -L+1 + \ord(\epsilon_3) \Big]
\\[3 ex]
C_o^{(4)}\Big[-\frac{1}{24 \epsilon_4 } -\frac{L}{6} + \frac{11-6 \ln 2}{36}    + \ord(\epsilon_4) \Big]
 \end{array}
\right.
\nn\\[4 ex]
\lefteqn{\raisebox{-0.5 ex}{\includegraphics[width=0.2 \textwidth]{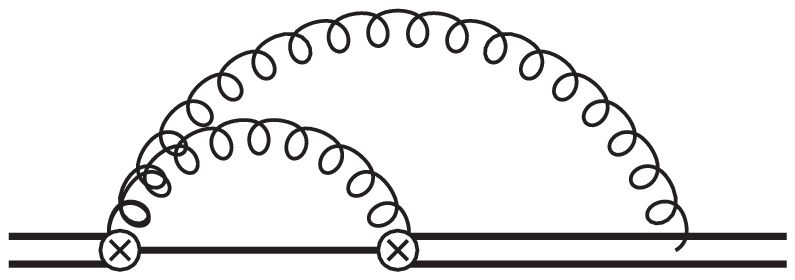} } = C_o^{(d)}\, 
\frac{(d-2) \Gamma (5-2 d) \Gamma \left(\frac{d}{2}-1\right)^2}{8 (d-3)} } & \\
&=\left\{ \begin{array}{l}
C_o^{(3)}\Big[-\frac{1}{4 \epsilon_3 ^2} + \frac{1-2 L}{2 \epsilon_3 } -2 L^2 +2 L -\frac{11 \pi ^2}{24}-2 + \ord(\epsilon_3) \Big]
\\[3 ex]
C_o^{(4)}\Big[-\frac{1}{6 \epsilon_4 } -\frac{2 L}{3} +\frac{25}{18}-\frac{2 \ln 2}{3}  + \ord(\epsilon_4) \Big]
 \end{array}
\right.
\nn\\[4 ex]
\lefteqn{\raisebox{-0.5 ex}{\includegraphics[width=0.2 \textwidth]{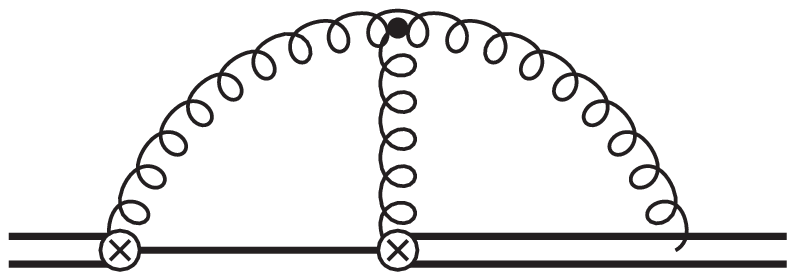} } = C_o^{(d)}\, \Bigg[
\frac{2^{1-2 d} \left(d^2-6 d+10\right) \Gamma \left(\frac{1}{2}-d\right) \Gamma (1-d)^2 \Gamma \left(d+\frac{1}{2}\right) \Gamma (d)^2}{(d-4) \Gamma
   \left(\frac{d-1}{2}\right) \Gamma \left(\frac{d+1}{2}\right)}} && \nn\\
&+\frac{\sqrt{\pi } 2^{-d-2} \left(-3 d^4+30 d^3-111 d^2+181 d-112\right) \Gamma (5-2 d) \Gamma
   \left(\frac{d}{2}-2\right) \Gamma (d-4)}{\Gamma \left(\frac{d+1}{2}\right) } \Bigg]
\\
&= \left\{ \begin{array}{l}
C_o^{(3)}\Big[-\frac{1}{2 \epsilon_3 } -2 L + \frac{1}{6} \left(15-\pi ^2\right)  + \ord(\epsilon_3) \Big]
\\[3 ex]
C_o^{(4)}\Big[-\frac{1}{8 \epsilon_4^2} + \frac{-9 L-2 \pi ^2+15-9 \ln 2}{18 \epsilon_4 } - L^2 - \frac{1}{9} L \left(-30+4 \pi ^2+ 18 \ln 2 \right)\\
\qquad + \frac{1}{6} (8 \zeta (3)-27+20 \ln 2 - 6 \ln^2 2) -\frac{\pi ^2}{432}  (192 \ln 2-47)  + \ord(\epsilon_4) \Big]
 \end{array}
\right.
\nn\displaybreak \\[4 ex]
\lefteqn{\raisebox{-0.5 ex}{\includegraphics[width=0.2 \textwidth]{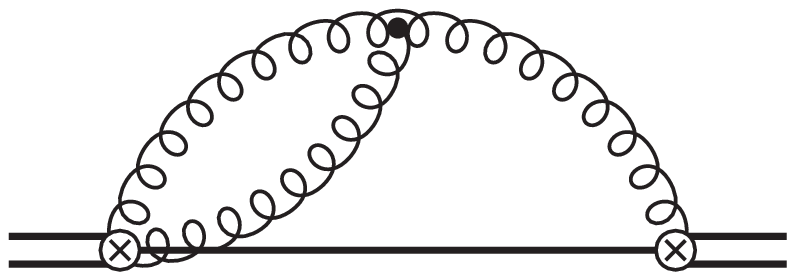} } = C_o^{(d)}\, 
\frac{3 (3-d) \Gamma (5-2 d) \Gamma \left(1-\frac{d}{2}\right) \Gamma \left(\frac{d}{2}-1\right) \Gamma \left(\frac{d}{2}\right)}{16 \Gamma
   \left(3-\frac{d}{2}\right)} }&&    \\
&=\left\{ \begin{array}{l}
C_o^{(3)}\Big[ -3  + \ord(\epsilon_3) \Big]
\\[3 ex]
C_o^{(4)}\Big[\frac{1}{8 \epsilon_4 ^2} +\frac{3 L-4+3 \ln 2}{6 \epsilon_4 } + L^2 +\frac{2L (3 \ln 2-4)}{3} +\frac{3 \pi ^2}{16}+\frac{26}{9}+\ln^2 2 -\frac{8 \ln 2}{3}  + \ord(\epsilon_4) \Big]
 \end{array}
\right.
\nn \\[4 ex]
\lefteqn{\raisebox{-0.5 ex}{\includegraphics[width=0.2 \textwidth]{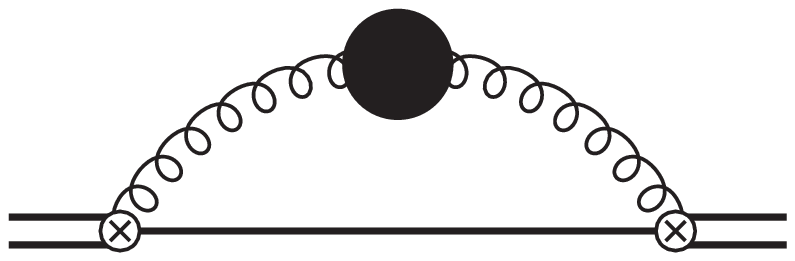} } = C_o^{(d)}  \frac{C_A (3 d-2)-4 (d-2) n_f T_F }{C_A } \times} \nn\\ 
&\hspace{30 ex} \times\frac{\sqrt{\pi } (d-3) \Gamma (5-2 d) \Gamma \left(1-\frac{d}{2}\right) \Gamma (d-2) \Gamma \left(\frac{d}{2}\right) }{2^{d+3}  \Gamma \left(3-\frac{d}{2}\right) \Gamma \left(\frac{d+1}{2}\right)}  &    \\
&=\left\{ \begin{array}{l}
C_o^{(3)}\Big[ \frac{7}{4}-\frac{n_f T_F}{C_A}  + \ord(\epsilon_3) \Big]
\\[3 ex]
C_o^{(4)} \Big[ 
\big[\frac{4 n_f T_F}{9 C_A} -\frac{5}{9}\big] \frac{1}{8 \epsilon_4^2}
+\frac{1}{2 \epsilon_4 }  \left( \big[\frac{4 n_f T_F}{9 C_A} -\frac{5}{9}\big] (L - \frac{5}{4} + \ln 2 ) + \frac{1}{18} \right)
+\frac{L}{9}
-\frac{1}{6}
+\frac{\ln 2}{9} \nn\\
\qquad + \big[\frac{4 n_f T_F}{9 C_A} -\frac{5}{9}\big] \left(L^2+L \left(2 \ln 2-\frac{5}{2}\right)+\frac{3 \pi ^2}{16}+\frac{95}{36}+\ln^2 2-\frac{5 \ln 2}{2}\right)
  + \ord(\epsilon_4) \Big]
 \end{array}
\right.
\nn
\end{align}

\begin{align}
L :=& \ln\Big[- \frac{\Delta V}{\nu_\msb}  \Big] \;;\quad  \nu_\msb := \nu\, e^{- \frac12(\gamma_E - \ln (4 \pi))}\\[2 ex]
C_o^{(d)} :=& 
-16 \,i\, C_A \Big(\frac{C_A}{2}-C_F\Big) \pi ^{2-d} r^2 V_A^2 \alpha ^2 (-\Delta V)^{2 d-5} \nu_{\msb} ^{-4 \epsilon_n } e^{-2 \epsilon_n  (\gamma_E
   -\ln (4 \pi ))} \\[2 ex]
C_o^{(3)} :=& -i\, C_A \Big(\frac{C_A}{2}-C_F \Big) \Delta V r^2 V_A^2 \alpha ^2 \\[2 ex]
C_o^{(4)} :=&  -\frac{i\, C_A \Big(\frac{C_A}{2}-C_F \Big) \Delta V^3 r^2 V_A^2 \alpha ^2}{\pi ^2} 
\end{align}

%\bea
%\text{``oct``-``sing`` (Eq.~\ref{Diag5})} &=& -2^{-2 d-1} (d-1) \Gamma \left(\frac{1}{2}-d\right) \Gamma \left(\frac{1}{2}-\frac{d}{2}\right)^2 %\Gamma \left(d+\frac{1}{2}\right) \\
%&=&\frac{\pi }{64 \epsilon_3 ^2} + \ord(\epsilon_3^{-1})\\
%&=&-\frac{\pi ^2}{96} + \ord(\epsilon_4)
%\eea
\subsection{Relation between singlet and octet diagrams}

\begin{align}
\raisebox{-0.5 ex}{\includegraphics[width=0.2 \textwidth]{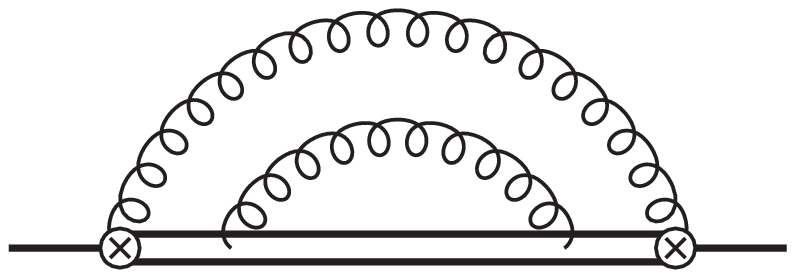} } & \xrightarrow[\Delta V \to - \Delta V]{C_F \to (\frac{1}{2}C_A-C_F)}  \raisebox{-0.5 ex}{\includegraphics[width=0.2 \textwidth]{Oct2loop_1.eps} }
\\[2 ex]
\raisebox{-2.8 ex}{\includegraphics[width=0.2 \textwidth]{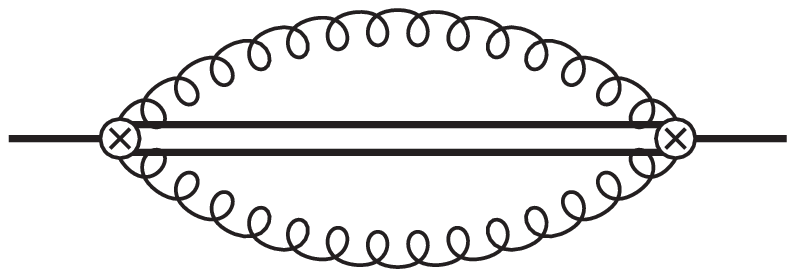} } & \xrightarrow[\Delta V \to - \Delta V]{C_F \to (\frac{1}{2}C_A-C_F)}  \raisebox{-2.8 ex}{\includegraphics[width=0.2 \textwidth]{Oct2loop_2.eps} }
\\[2 ex]
\raisebox{-0.5 ex}{\includegraphics[width=0.2 \textwidth]{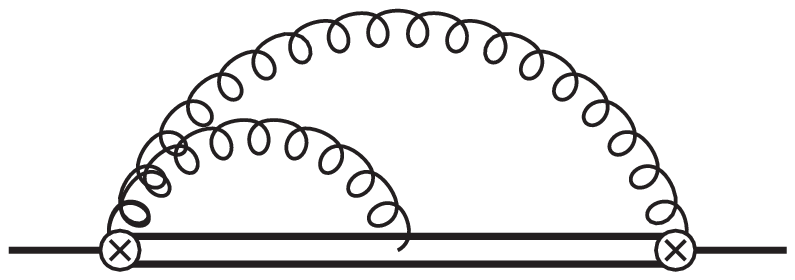} } & \xrightarrow[\Delta V \to - \Delta V]{C_F \to (\frac{1}{2}C_A-C_F)}  \raisebox{-0.5 ex}{\includegraphics[width=0.2 \textwidth]{Oct2loop_4.eps} }
\\[2 ex]
\raisebox{-0.5 ex}{\includegraphics[width=0.2 \textwidth]{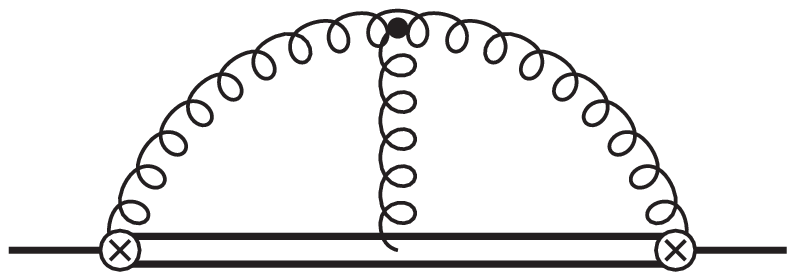} } & \xrightarrow[\Delta V \to - \Delta V]{C_F \to (\frac{1}{2}C_A-C_F)}  \raisebox{-0.5 ex}{\includegraphics[width=0.2 \textwidth]{Oct2loop_5.eps} } \lefteqn{+ \Delta_{so} }
\end{align}

\begin{align}
\Delta_{so} &= -C_o^{(d)}\, \frac{4^{-d} \left(d^2-6 d+10\right) \Gamma \left(\frac{1}{2}-d\right) \Gamma \left(\frac{1}{2}-\frac{d}{2}\right)^2 \Gamma
   \left(\frac{d}{2}+\frac{1}{2}\right)^2 \Gamma \left(d+\frac{1}{2}\right)}{(d-4) \Gamma \left(\frac{d-1}{2}\right) \Gamma \left(\frac{d+1}{2}\right)}
=
\\
&=
\left\{ \begin{array}{l}
C_o^{(3)}\Big[ \frac{1}{4 \epsilon_3^2} + \frac{4 L+1}{4 \epsilon_3 } + 2 L^2 + L + \frac{13 \pi ^2}{24}+\frac{7}{4}  + \ord(\epsilon_3) \Big]
\\[3 ex]
C_o^{(4)}\Big[ \frac{\pi ^2}{6 \epsilon_4 } + \frac{2 \pi ^2 L}{3} + \frac{2}{9} \pi ^2 (3\ln 2 -2)  + \ord(\epsilon_4) \Big]
 \end{array}
\right.
\end{align}

\begin{align}
\raisebox{-0.5 ex}{\includegraphics[width=0.2 \textwidth]{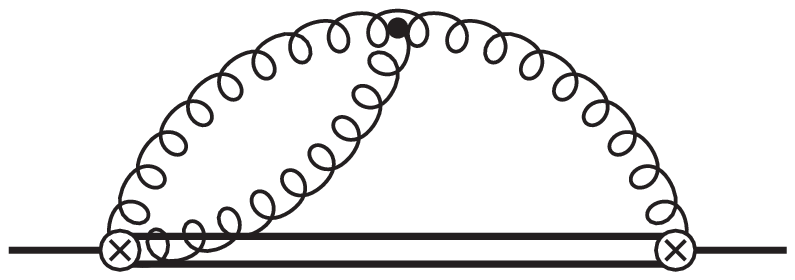} } & \xrightarrow[\Delta V \to - \Delta V]{C_F \to (\frac{1}{2}C_A-C_F)}  \raisebox{-0.5 ex}{\includegraphics[width=0.2 \textwidth]{Oct2loop_6.eps} }
\\[2 ex]
\raisebox{-0.5 ex}{\includegraphics[width=0.2 \textwidth]{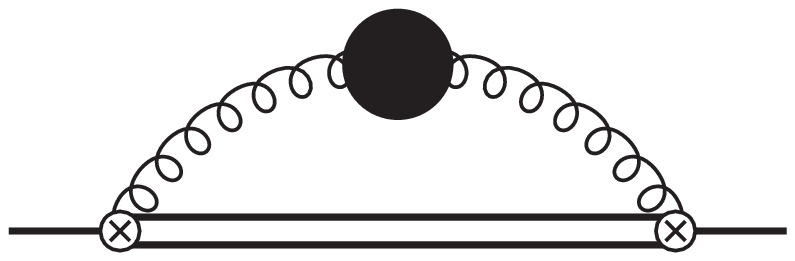} } & \xrightarrow[\Delta V \to - \Delta V]{C_F \to (\frac{1}{2}C_A-C_F)} \raisebox{-0.5 ex}{\includegraphics[width=0.2 \textwidth]{Oct2loop_7.eps} }
\end{align}


\begin{thebibliography}{}

%\cite{Pineda:2010mb}
\bibitem{Pineda:2010mb}
  A.~Pineda and M.~Stahlhofen,
  %``The QCD static potential in D<4 dimensions at weak coupling,''
  Phys.\ Rev.\  D {\bf 81}, 074026 (2010)
  [arXiv:1002.1965 [hep-th]].
  %%CITATION = PHRVA,D81,074026;%%

%\cite{Jorysz:1987qj}
\bibitem{Jorysz:1987qj}
  I.~H.~Jorysz and C.~Michael,
  %``THE FIELD CONFIGURATIONS OF A STATIC ADJOINT SOURCE IN SU(2) LATTICE GAUGE
  %THEORY,''
  Nucl.\ Phys.\  B {\bf 302} (1988) 448.
  %%CITATION = NUPHA,B302,448;%%
  
\bibitem{FM}
M.~Foster and C.~Michael  [UKQCD Collaboration],
%``Hadrons with a heavy colour-adjoint particle,''
Phys.\ Rev.\ D {\bf 59}, 094509 (1999)
[arXiv:hep-lat/9811010].
%%CITATION = HEP-LAT 9811010;%%

  %\cite{Brambilla:1999xf}
\bibitem{Brambilla:1999xf}
  N.~Brambilla, A.~Pineda, J.~Soto and A.~Vairo,
  %``Potential NRQCD: An effective theory for heavy quarkonium,''
  Nucl.\ Phys.\  B {\bf 566}, 275 (2000)
  [arXiv:hep-ph/9907240].
  %%CITATION = NUPHA,B566,275;%%
  
%\cite{Bali:2003jq}
\bibitem{Bali:2003jq}
  G.~S.~Bali and A.~Pineda,
  %``QCD phenomenology of static sources and gluonic excitations at short
  %distances,''
  Phys.\ Rev.\  D {\bf 69}, 094001 (2004)
  [arXiv:hep-ph/0310130].
  %%CITATION = PHRVA,D69,094001;%%
  
%\cite{Pineda:1997bj}
\bibitem{Pineda:1997bj}
  A.~Pineda and J.~Soto,
  %``Effective field theory for ultrasoft momenta in NRQCD and NRQED,''
  Nucl.\ Phys.\ Proc.\ Suppl.\  {\bf 64}, 428 (1998).
 % [arXiv:hep-ph/9707481].
  %%CITATION = NUPHZ,64,428;%%

%\cite{Brambilla:2004jw}
\bibitem{Brambilla:2004jw}
  N.~Brambilla, A.~Pineda, J.~Soto and A.~Vairo,
  %``Effective field theories for heavy quarkonium,''
  Rev.\ Mod.\ Phys.\  {\bf 77}, 1423 (2005).
%  [arXiv:hep-ph/0410047].
  %%CITATION = RMPHA,77,1423;%%

%\cite{Kniehl:2004rk}
\bibitem{Kniehl:2004rk}
  B.~A.~Kniehl, A.~A.~Penin, Y.~Schroder, V.~A.~Smirnov and M.~Steinhauser,
  %``Two-loop static QCD potential for general colour state,''
  Phys.\ Lett.\  B {\bf 607}, 96 (2005)
  [arXiv:hep-ph/0412083].
  %%CITATION = PHLTA,B607,96;%%

\bibitem{short} 
  N.~Brambilla, A.~Pineda, J.~Soto and A.~Vairo,
  %``The infrared behaviour of the static potential in perturbative {QCD},''
  Phys.\ Rev.\  D {\bf 60}, 091502 (1999).
%  [arXiv:hep-ph/9903355].
  %%CITATION = PHRVA,D60,091502;%%
    

\bibitem{RG} 
  A.~Pineda and J.~Soto,
  %``The renormalization group improvement of the QCD static potentials,''
  Phys.\ Lett.\  B {\bf 495}, 323 (2000).
%  [arXiv:hep-ph/0007197].
  %%CITATION = PHLTA,B495,323;%%

%\cite{Brambilla:2010xn}
\bibitem{Brambilla:2010xn}
  N.~Brambilla, J.~Ghiglieri, P.~Petreczky and A.~Vairo,
  %``Polyakov loop and correlator of Polyakov loops at next-to-next-to-leading
  %order,''
  Phys.\ Rev.\  D {\bf 82}, 074019 (2010)
  [arXiv:1007.5172 [hep-ph]].
  %%CITATION = PHRVA,D82,074019;%%

%\cite{Eidemuller:1997bb}
\bibitem{Eidemuller:1997bb}
  M.~Eidemuller and M.~Jamin,
  %``QCD field strength correlator at the next-to-leading order,''
  Phys.\ Lett.\  B {\bf 416}, 415 (1998)
  [arXiv:hep-ph/9709419].
  %%CITATION = PHLTA,B416,415;%%

%\cite{Brambilla:2006wp}
\bibitem{Brambilla:2006wp}
  N.~Brambilla, X.~Garcia i Tormo, J.~Soto and A.~Vairo,
  %``The logarithmic contribution to the QCD static energy at NNNNLO,''
  Phys.\ Lett.\  B {\bf 647}, 185 (2007).
%  [arXiv:hep-ph/0610143].
  %%CITATION = PHLTA,B647,185;%%

%\cite{Brambilla:2009bi}
\bibitem{Brambilla:2009bi}
  N.~Brambilla, A.~Vairo, X.~Garcia i Tormo and J.~Soto,
  %``The QCD static energy at NNNLL,''
  Phys.\ Rev.\  D {\bf 80}, 034016 (2009)
  [arXiv:0906.1390 [hep-ph]].
  %%CITATION = PHRVA,D80,034016;%%

 %\cite{Schroder:1999sg}
\bibitem{Schroder:1999sg}
  Y.~Schroder,
  ``The static potential in QCD'', DESY-THESIS-1999-021.
  %%CITATION = DESY-THESIS-1999-021;%%

%\cite{Schroder:1998vy}
\bibitem{Schroder:1998vy}
  Y.~Schroder,
  %``The Static potential in QCD to two loops,''
  Phys.\ Lett.\  B {\bf 447}, 321 (1999)
  [arXiv:hep-ph/9812205].
  %%CITATION = PHLTA,B447,321;%%

%\cite{Pascual:1984zb}
\bibitem{Pascual:1984zb}
  P.~Pascual and R.~Tarrach,
  %``QCD: Renormalization For The Practitioner,''
  Lect.\ Notes Phys.\  {\bf 194}, 1 (1984).
  %%CITATION = LNPHA,194,1;%%

%\cite{Pineda:1997ie}
\bibitem{Pineda:1997ie}
  A.~Pineda and J.~Soto,
  %``The Lamb Shift in Dimensional Regularization,''
  Phys.\ Lett.\  B {\bf 420}, 391 (1998)
  [arXiv:hep-ph/9711292].
  %%CITATION = PHLTA,B420,391;%%

\bibitem{KP1} 
  B.~A.~Kniehl and A.~A.~Penin,
  %``Ultrasoft effects in heavy quarkonium physics,''
  Nucl.\ Phys.\  B {\bf 563}, 200 (1999).
%  [arXiv:hep-ph/9907489].
  %%CITATION = NUPHA,B563,200;%%

%\cite{Hand:1995rg}
\bibitem{Hand:1995rg}
  B.~J.~Hand and G.~Leibbrandt,
  %``The Wilson Loop in Yang-Mills Theory in the General Axial Gauge,''
  arXiv:hep-th/9511140.
  %%CITATION = HEP-TH/9511140;%%

%\cite{Kniehl:1998fn}
\bibitem{Kniehl:1998fn}
  B.~A.~Kniehl and A.~Sirlin,
  %``Differences between the pole and on-shell masses and widths of the  Higgs
  %boson,''
  Phys.\ Rev.\ Lett.\  {\bf 81}, 1373 (1998)
  [arXiv:hep-ph/9805390].
  %%CITATION = PRLTA,81,1373;%%  

 \bibitem{FSP} W. Fischler, Nucl. Phys. {\bf B129}, 157 (1977).

%\cite{Philipsen:2002az}
\bibitem{Philipsen:2002az}
  O.~Philipsen,
  %``Nonperturbative formulation of the static color octet potential,''
  Phys.\ Lett.\  B {\bf 535}, 138 (2002)
  [arXiv:hep-lat/0203018].
  %%CITATION = PHLTA,B535,138;%%

%\cite{Juge:2003ge}
\bibitem{Juge:2003ge}
  K.~J.~Juge, J.~Kuti and C.~Morningstar,
  %``Excitations of the static quark anti-quark system in several gauge
  %theories,''
  arXiv:hep-lat/0312019.
  %%CITATION = HEP-LAT/0312019;%%

%\cite{Caselle:2004er}
\bibitem{Caselle:2004er}
  M.~Caselle, M.~Pepe and A.~Rago,
  %``Static quark potential and effective string corrections in the (2+1)-d
  %SU(2) Yang-Mills theory,''
  JHEP {\bf 0410}, 005 (2004)
  [arXiv:hep-lat/0406008].
  %%CITATION = JHEPA,0410,005;%%

%\cite{Kuti:2005xg}
\bibitem{Kuti:2005xg}
  J.~Kuti,
  %``Lattice QCD and string theory,''
  PoS {\bf LAT2005}, 001 (2006)
  [PoS {\bf JHW2005}, 009 (2006)]
  [arXiv:hep-lat/0511023].
  %%CITATION = POSCI,JHW2005,009;%%

%\cite{Brandt:2009tc}
\bibitem{Brandt:2009tc}
  B.~B.~Brandt and P.~Majumdar,
  %``Spectrum of the QCD flux tube in 3d SU(2) lattice gauge theory,''
  Phys.\ Lett.\  B {\bf 682}, 253 (2009)
  [arXiv:0905.4195 [hep-lat]].
  %%CITATION = PHLTA,B682,253;%%

%\cite{Binosi:2003yf}
\bibitem{JaxoDraw}
  D.~Binosi and L.~Theussl,
  %``JaxoDraw: A graphical user interface for drawing Feynman diagrams,''
  Comput.\ Phys.\ Commun.\  {\bf 161}, 76 (2004)
  [arXiv:hep-ph/0309015].
  %%CITATION = CPHCB,161,76;%%


\end{thebibliography}
\end{document}